\documentclass[%
reprint,
superscriptaddress,
showpacs,preprintnumbers,
nofootinbib,
amsmath,amssymb,
aps,
pra,
showkeys,
floatfix,
twocolumn,
]{revtex4-1}
\usepackage{graphicx}
\usepackage[caption=false]{subfig}
\usepackage{dcolumn}
\usepackage{bm}
\usepackage{dsfont}
\usepackage{amsmath,amssymb,stackengine}
\usepackage{siunitx}
\usepackage[caption=false]{subfig}
\usepackage{braket}
\usepackage{listings}
\usepackage{xcolor}
\usepackage{hyperref}
\hypersetup{colorlinks=true,linkcolor=blue,urlcolor=blue,citecolor=blue}
\begin{document}
\title{Pancharatnam-Berry phase in neutrino mixing}
\author{Manosh T.M.}
\email{tm.manosh@cusat.ac.in}
\email{tm.manosh@gmail.com}
\affiliation{Department of Physics, Cochin University of Science and Technology, Kochi 682--022, India.}
\author{N. Shaji}
\email{shajin@cusat.ac.in}
\affiliation{Department of Physics, Cochin University of Science and Technology, Kochi 682--022, India.}
\author{Ramesh Babu Thayyullathil}
\email{rbt@cusat.ac.in}
\author{Titus K. Mathew}
\email{titus@cusat.ac.in}
\affiliation{Department of Physics, Cochin University of Science and Technology, Kochi 682--022, India.}
\affiliation{Centre for Particle Physics, Cochin University of Science and Technology, Kochi 682--022, India.}
\date{\today}

\begin{abstract}
The Pancharatnam - Berry phase (PBP) of purely geometric origin appears as a reparametrization invariant quantity of ray Space. In this article, we investigate the properties exhibited by PBP in neutrino mixing. We map the neutrino flavour modes to independent flavour vacuum states and compute PBP using Bargmann invariant. We derive the exact formula for PBP in two flavour approximation using the kinematic approach. Our result reproduces previous results of Blasone \textit{et al.} \cite{BLASONE1999262} under cyclic condition. Inspired by the work of Mukunda and Simon \cite{MUKUNDA1993205}, we investigate the total and dynamical phases separately. This method leads us to identify the existence of nodal points in the mixing parameter space. At nodal points, PBP changes by a value $\pi$, and it originates from the total phase. We report the direct relation between nodal points and MSW resonance, giving physical meaning to nodal points. Our analysis shows the ability of PBP to differentiate between different mass hierarchies and set numerical bounds to $\Delta m^2$ by changing total energy.  We extend our studies to three flavour model and found that PBP is sensitive to the Dirac $CP$ phase ($\delta_{CP}$). Using our $N$-qubit architecture of the $N$-flavour neutrino model, one can immediately study the dynamical characteristics like mode entanglement between neutrino flavour modes.
\end{abstract}
\keywords{Pancharatnam-Berry phase, neutrinos, nodal points}
\maketitle
\section{Introduction}

The physics of mixing attract a diverse spectrum of researchers to work on neutrinos. Studies range from the measurement of mixing parameters using neutrino oscillation to the demonstration of practical communication channels using neutrino beam \cite{PhysRevLett.123.151803, doi:10.1142/S0217732312500770}. Apart from the standard vacuum oscillation, due to the mixing of mass and flavour eigenstates, exciting works ranging from probing planetary and stellar interiors \cite{PhysRevD.96.036005, Rott2015, PhysRevD.56.8082} to the effect of gravitational coupling on the Mikheyev-Smirnov-Wolfenstein (MSW) effect \cite{PhysRevLett.67.1833} also exist.

There were motivations to study neutrino - antineutrino oscillation based on kaon mixing even before having evidence for different neutrino flavours. Later, with the discovery of different flavour species, the celebrated Pontecorvo-Maki-Nakagawa-Sakata (PMNS) formalism got broad appreciation. Blasone \textit{et al.} \cite{PhysRevD.69.057301, PhysRevD.66.025033}  derived the exact formulas of mixing and oscillations of neutral boson, Majorana and Dirac fields in the framework of quantum field theory (QFT). The QFT description of neutrino flavour mixing adds corrections to the existing quantum mechanical expressions and allows one to investigate more features like topology \cite{BLASONE1999262} and quantum resources \cite{Blasone_2009, Blasone2014}.  In the ultra-relativistic limit, where the neutrino mass is significantly less than the total energy, the QFT representation reduces to the standard PMNS formalism. Hence, one can describe flavour modes as excitations of the flavour Fock states \cite{Blasone_2009}. Similar to the two-level system in quantum optics \cite{doi:10.1142/p380} one can readily map the flavour modes to individual qubits and describe the dynamics with a flavour Hamiltonian. Recently, the article by Derby \textit{et al.} \cite{derby2021compact}, and the references within, demonstrated a novel method to map fermions to qubits, which could be of interest for many body simulations with quantum computers.

In neutrinos, the non-degenerate mass eigenstates are responsible for mixing and flavour oscillation. Analogous to the phase difference generated by the path length in the optical double-slit experiments, the mass difference produces a phase proportional to the mass squared differences and leads to interference. Giunti \cite{Giunti_2003} gave a comprehensive description of this phase factor by meticulously studying the phase velocities. Apart from the path difference, Pancharatnam \cite{Pancharatnam1956} discovered a noticeable change in the interference pattern when a polarized light beam undergoes a rotation. This phase is identical to the phase factor in quantum adiabatic processes discovered by Berry \cite{doi:10.1098/rspa.1984.0023}. This phase factor, widely known as the Pancharatnam-Berry phase (called PBP hereafter), goes by different names and appears in diverse areas of physics. Conventionally, Hamiltonians with slowly varying external parameters can generate PBP. In the case of neutrinos, a slowly varying matter density can make the Hamiltonian parameter dependent. Naumov \cite{NAUMOV1994351} investigated this and concluded that three flavour mixing is essential for neutrinos to show PBP. Later, Blasone \textit{et al.} \cite{BLASONE1999262} proved that a time-dependent canonical transformation gives a gauge like structure to the Hamiltonian. Then, PBP could arise naturally in neutrino oscillation even for two flavour scenario. Consequently, most authors take two distinct routes in analyzing PBP in neutrinos. One is to investigate the parametric dependence of the Hamiltonian \cite{PhysRevD.72.053012,JOSHI2016135,PhysRevD.96.096004} and the other explores the natural gauge structure arising from the time evolution \cite{Mukunda1997,JOSHI2020135766,CAPOLUPO2018216,BLASONE200973,PhysRevD.63.053003,Dixit_2018}. Articles by Mehta \cite{PhysRevD.79.096013} and Johns \textit{et al.} \cite{PhysRevD.95.043003} give an unbiased, thorough picture of the geometric phases in neutrino physics. 

This article considers neutrino flavour modes as excitations of independent Fock states and computes the composite PBP for various mixing parameters. Our analysis treats both routes mentioned before in a unified manner and gives a holistic picture of PBP in the kinematic approach. We highlight the results produced by other authors and explains the underlying reasons for such conclusions. We identify the existence of nodal points and give physical interpretation based on MSW resonance. We examine the ability of PBP to reflect the effects produced by matter potential, different mass hierarchies and the Dirac $CP$ phase. 

The paper is structured as follows. In Sec. II, we give a brief introduction to the PBP relevant to our discussion. Later, in Sec. III, we construct the flavour Hamiltonian and give explicit representation in $2^N$ dimensional Hilbert space and describe the unitary time evolution operator. In Sec. IV, we derive the exact formula for PBP with two flavour approximation and give an efficient numerical method to compute PBP for three flavour scenario. Further, we explore the effects of matter potential, mass hierarchies and the Dirac $CP$ phases on PBP and discuss the importance of nodal points under the same section. Finally, we conclude in Sec V.

\section{Pancharatnam-Berry phase (PBP)}
Pancharatnam-Berry phase (PBP) is extensively investigated in classical and quantum optics with many debates and discussions. In the seminal work of Fuentes-Guridi \textit{et al.} \cite{PhysRevLett.89.220404}, it is shown how vacuum can induce PBP to a spin half system by mixing the field modes. However, Larson \cite{PhysRevLett.108.033601} proposed that such PBP is an artefact of rotating wave approximation (RWA). Later, Calder\'{o}n \textit{et al.} \cite{PhysRevA.93.033823} presented concrete evidence for vacuum induced PBP with and without RWA.  Finally, Gasparinetti \cite{Gasparinettie1501732} gave solid experimental evidence for vacuum induced PBP, confirming the fourteen-year-old prediction made by Fuentes-Guridi \textit{et al.}.

A major breakthrough in the field came with the work of Mukunda and Simon \cite{MUKUNDA1993205}, where they gave a quantum kinematic description to PBP. In the kinematic approach, state vector after a time $T$ acquires a total phase such that, $\ket{\psi(T)}=\text{e}^{i\phi}\ket{\psi(0)}$. The total phase ($\phi$) is the sum of the dynamical ($-\Theta$)\footnote{Not to be confused with the mixing angle, $\theta$, in PMNS matrix.} and the geometrical ($\Phi$) phases. After time $T$, the dynamical phase is given by,
\begin{equation}
-\Theta=-i\int_{0}^{T}\bra{\psi(t)}\partial_t\ket{\psi(t)}dt. 
\label{Eq:1}
\end{equation} 
Then, by definition, the geometric phase takes the form,
\begin{equation}
\Phi=\phi+i\int_{0}^{T}\bra{\psi(t)}\partial_t\ket{\psi(t)}dt.
\label{Eq:2}
\end{equation}
Expressing the above equation as argument of exponentials, we get, 
\begin{align}
&\Phi=\arg\left[\exp({i\phi})\exp\left({-\int_{0}^{T}\bra{\psi(t)}\partial_t\ket{\psi(t)}dt}\right)\right]\nonumber\\&=\arg\left[\braket{\psi(0)|\psi(T)}\exp\left(-\int_{0}^{T}\bra{\psi(t)}\partial_t\ket{\psi(t)}dt\right)\right].
\label{Eq:3}
\end{align}
Where, $\braket{\psi(0)|\psi(T)}$ is the total phase and the rest accounts the dynamical phase. The minus sign in the dynamical phase is due to the convention followed in the Schr\"{o}dinger equation. Hence, PBP is a real valued, re-parametrization invariant quantity of ray space. Sj\"{o}qvist \cite{PhysRevA.62.022109} identified that this expression is also valid for non-cyclic unitary evolution. Further Sj\"{o}qvist \textit{et al.} \cite{PhysRevLett.85.2845} generalized the expression to mixed states. Later, Filipp \textit{et al.} \cite{PhysRevLett.90.050403} introduced the idea of off-diagonal PBP for mixed states and highlighted the features of nodal points. Finally, Tong \textit{et al.} \cite{PhysRevLett.93.080405} generalized the Eq. (\ref{Eq:3}) to mixed states undergoing non-unitary evolution as, 
\begin{align}
\Phi=&\arg\left[\sum_{j}\sqrt{\epsilon_j(0)\epsilon_j(T)}\braket{j(0)|j(T)}\right.\nonumber\\
&\times\left.\exp\left(-\int_{0}^{T}\bra{j(t)}\partial_t\ket{j(t)}dt\right)\right].
\label{Eq:4}
\end{align}
Here, $\ket{j(t)}$ and $\epsilon_j(t)$ corresponds to the $j^{\text{th}}$ eigenstate and eigenvalue of the density matrix $\rho(t)$ at time `$t$' respectively. For pure states, Eq. (\ref{Eq:4}) reduces to Eq. (\ref{Eq:3}). For simplicity, we restrict ourself to pure states and calculate PBP using Eq. (\ref{Eq:3}). The kinematic approach offers a simplified calculation of PBP as a transformation invariant quantity. Motivated from the condition that, the PBP about a geodesic in ray space is zero, Mukunda, \cite{Mukunda1997} identified the first order approximation of PBP as the Bargmann invariant. This implies, one can approximate the Eq. (\ref{Eq:3}) as,
\begin{align}
\Phi\approx&\arg\left[\braket{\psi(0)|\psi(T)}\braket{\psi(T)|\psi(T-\delta t)}\cdots\right.\nonumber\\
&\left.\cdots\times\braket{\psi(t_i)|\psi(t_{i-1})}\cdots\braket{\psi(\delta t)|\psi(0)}\right].
\label{Eq:5}
\end{align}
In the above expression, the first inner product accounts for the total phase and the following products account for the first-order approximation of the integral in Eq. (\ref{Eq:3}). Guo \textit{et al.} \cite{PhysRevA.90.062133} used this expression to analyze the non-Markovian dynamics of the dissipative qubit and found the existence of nodal points in PBP without RWA, an important point which was missed by Chen \textit{et al.} \cite{PhysRevA.81.022120}. In our work, we use Eq. (\ref{Eq:3}) for the explicit calculation of PBP. However, when analytical solutions become cumbersome, we use Eq. (\ref{Eq:5}).

For more generalized views on PBP, one may refer to Martinez \cite{PhysRevD.42.722}, in which the author generalizes PBP to field theory and Hsin \textit{et al.} \cite{PhysRevB.102.245113} where the authors investigate the diabolical points and boundary phenomena associated with PBP in QFT.  A recent review by Jisha \textit{et al.} \cite{JishaCP2021} (private communication) gives a solid description of PBP in optics and its applications.

\section{Neutrino flavour modes}

Balsone \textit{et al.} \cite{PhysRevD.66.025033} gave the exact formulas for neutrino oscillation by considering a QFT description of fermion mixing. In the QFT description, the flavour states along with the creation and annihilation operators obeys a deformed $su(3)$ algebra. In all practical cases neutrinos are ultra-relativistic; i.e. the momentum of neutrinos are much larger than the actual neutrino mass, and one can approximate neutrino momentum to its total energy. Now, considering a plane wave solution, we can assume the energy difference between the mass eigenstates as, $\Delta E_{ij}=E_i-E_j\approx\Delta m^2_{ij}/(2E)$. Where, $\Delta m^2_{ij}=m_i^2-m_j^2$ is the mass squared difference and $E$ is the total energy. The flavour eigenstates in the QFT description of neutrino mixing can be taken as the flavour eigenstates in the PMNS formalism in the ultra-relativistic limit. This framework allows one to write the flavour eigenstates as single-mode excitations of flavour vacuum states.  With, $\ket{\nu_e}=\ket{1}\otimes\ket{0}\otimes\ket{0}=\ket{100}$, $\ket{\nu_{\mu}}=\ket{0}\otimes\ket{1}\otimes\ket{0}=\ket{010}$ and $\ket{\nu_{\tau}}=\ket{0}\otimes\ket{0}\otimes\ket{1}=\ket{001}$, the time evolution of any given flavour (say $\alpha$) takes the form,
\begin{equation}
\ket{\nu_{\alpha}(t)}=\mathds{U}_{\alpha e}(t)\ket{100}+\mathds{U}_{\alpha\mu}(t)\ket{010}+\mathds{U}_{\alpha\tau}(t)\ket{001}.
\label{Eq:6}
\end{equation}
Here,
\begin{equation}
\ket{0}=\begin{pmatrix}
1\\0
\end{pmatrix}\,\text{and}\,\ket{1}=\begin{pmatrix}
0\\1
\end{pmatrix}.
\label{Eq:7}
\end{equation}
This architecture is very similar to the W state observed in quantum information theory. Hence, one can explore traits like single-particle entanglement and other quantum discords from an information perspective. The time-dependent coefficients in the Eq. (\ref{Eq:6}) will be discussed later.

In general, Hamiltonian has no role in the kinematic approach. Since we are interested in PBP due to the circuit made by $\{\ket{\psi(t)}\}$, we solve the Schr\"{o}dinger equation to obtain the state vectors. For our purpose, we construct a flavour Hamiltonian in natural units. A plane wave approximation of mass eigenstates allows us to write the free Hamiltonian on an energy basis. Without the loss of generality, we can eliminate an overall phase factor of the form, $\mathds{1}m_1^2$, and reconstruct the Hamiltonian in terms of $\Delta m^2_{ij}=m_i^2-m_j^2$. One can then write,
\begin{equation}
\hat{H}_{m^2}=\text{diag}\left(0,\Delta m_{31}^2,\Delta m_{21}^2,0,0,0,0,0\right).
\label{Eq:8}
\end{equation}
The above construction follows directly from the basis used in Eq. (\ref{Eq:6}). The factor $1/E$ will be included later for convenience. In short, we have mapped a three-level quantum system to three two-level systems. This mapping might look like the overuse of qubits,  but mapping each mode into individual qubits allows one to explore quantum correlations among each mode.

Neutrinos are detected in flavour modes and the flavour modes are connected to the mass eigenstates via a unitary rotation called the PMNS matrix ($U$). Thus, we can write the Hamiltonian in flavour basis as,
\begin{equation}
\hat{H}_f=\frac{1}{2E}\left(U\hat{H}_{m^2}U^{\dagger}\pm\hat{V}\right).
\label{Eq:9}
\end{equation}
Here, $\hat{V}$ is the potential responsible for the coherent forward scattering of electron neutrinos and antineutrinos by matter. The $\pm$ sign corresponds to neutrinos and antineutrinos, respectively. One can also modify the elements of $\hat{V}$ to include non-standard interactions. The PMNS matrix $U$ is parametrized by three mixing angles, $\{\theta_{12},\theta_{13},\theta_{23}\}$ and the Dirac phase ($\delta_{CP}$). One can include the Majorana phases to the same, but an overall global phase does not contribute to the oscillation probabilities. Even though such phases are relevant for PBP discussions, for simplicity, we will consider them elsewhere. 

Given the Hamiltonian, one can solve the Schr\"{o}dinger equation to compute $\{\ket{\psi(t)}\}$. Since our Hamiltonian is Hermitian and time-independent, the unitary time evolution (with $\hbar=c=1$) is given by,
\begin{equation}
\mathds{U}(t)=\exp\left(-i\int_{0}^{t}\hat{H}_fdt'\right)=\exp\left(-i\hat{H}_ft\right).
\label{Eq:10}
\end{equation}
Evaluating $\mathds{U}(t)$ can be tedious depending on the form of $\hat{H}_f$. In our case, we map our system to a collection of three qubits, and hence, the Hamiltonian lives in a Hilbert space of the form $2\otimes2\otimes2$. We can decompose any Hermitian Hamiltonian of the from $2^N\times2^N$ into the linear combination of tensor products of Pauli matrices including identity ( $\left\{\sigma_1=\mathds{1},\sigma_x,\sigma_y,\sigma_z\right\}$) using the Hilbert-Schmidt decomposition given by,
\begin{equation}
\hat{H}=\sum_{i,j,\cdots,l}a_{ij\cdots l}\{\sigma_i\otimes\sigma_j\otimes\cdots\otimes\sigma_l\},
\label{Eq:11}
\end{equation}
with,
\begin{equation}
a_{ij\cdots l}=\frac{1}{2^N}\text{Tr}\left\{\left(\sigma_i\otimes\sigma_j\otimes\cdots\otimes\sigma_l\right)^{\dagger}\cdot \hat{H}\right\}.
\label{Eq:12}
\end{equation}
Here, $i,j,\cdots,l$ can take values $1,x,y$ and $z$. Thus, our Hamiltonian takes the form, $\hat{H}=\sum_{j}\hat{H}_j$. In general, each tensor products in the linear expression may not commute. This could arise from factors like detuning, external driving fields etc. Now, for non commuting $\{\hat{H}_j\}$, we express the unitary time evolution as the Lie-Trotter product, given by, 
\begin{equation}
\exp\left(-i\sum_j\hat{H}_jt\right)=\lim\limits_{M\rightarrow\infty}\left(\prod_{j}\exp\left(\frac{-i\hat{H}_jt}{M}\right)\right)^M.
\label{Eq:13}
\end{equation}
Lie-Trotter product is highly useful in implementing Hamiltonians in a quantum circuit. A recent article by Childs \textit{et al.} \cite{PhysRevX.11.011020} gives a tighter error bound by considering the commutativity of operators. When analytical methods become cumbersome, we use appropriate numerical methods \cite{JOHANSSON20131234} to solve the Schr\"{o}dinger equation.

Since we can take neutrinos to be ultra-relativistic, the approximation $t\approx L$ holds. Where, $L$ is the distance travelled, and is described as the baseline of oscillation experiments. Thus Eq. (\ref{Eq:10}) takes the from, $\mathds{U}(t)\approx\mathds{U}(L)=\exp\left(-i\hat{H}_fL\right)$. Further, it is convenient to deal the equations in natural units. For that, we represent $\Delta m^2_{ij}$ in eV$^2$, $L$ in km and $E$ in GeV. In this convention, the flavour Hamiltonian becomes,
\begin{equation}
\hat{H}_f=\frac{1.27\times2}{E}\left(U\hat{H}_{m^2}U^{\dagger}\pm\hat{V}\right).
\label{Eq:14}
\end{equation}
\subsection{Two flavour approximation}
Since the experimental evidence suggests a smaller value for $\theta_{13}$, a two flavour approximation is often valid for discussions on neutrino mixing. In the two flavour approximation, the general expression for a time dependent state vector is
\begin{equation}
\ket{\nu_{\gamma}(t)}=\mathds{U}_{\gamma \alpha}(t)\ket{10}+\mathds{U}_{\gamma\beta}(t)\ket{01}.
\label{Eq:15}
\end{equation}
Here, $\gamma\in\{\alpha,\beta\}$ and $\{\alpha, \beta\}$ can be any two combinations of lepton flavours. Since we are following the convention where, $\ket{\nu_e}=\ket{100}$, we shall rewrite $U$ and $\hat{H}_{m^2}$ accordingly. The choice of the convention is to match the literature by Blasone \textit{et al.}. In this convention, the PMNS mixing matrix is parameterized by a single mixing angle and is given by,
\begin{equation}
U=\begin{pmatrix}1 & 0 & 0 & 0\\0 & \cos{\theta} & \,\!\!\!-\sin{\theta} & 0\\0 & \sin{\theta} & \,\,\,\cos{\theta} & 0\\0 & 0 & 0 & 1\end{pmatrix},\label{Eq:16}.
\end{equation}
Under two flavour approximation, Eq. (\ref{Eq:8}) takes the from,
\begin{equation}
\hat{H}_{m^2}=\text{diad}\left(0,\Delta{m^2},0,0\right).
\label{Eq:17}
\end{equation}
The number indices are dropped as there is only one $\Delta{m^2}$. 

For neutrinos propagating in vacuum we can set $\hat{V}=0$ in Eq. (\ref{Eq:14}) and $\hat{H}_f$ becomes,
\begin{align}
&\hat{H}_f=\frac{1.27\times2}{E}\left(U\hat{H}_{m^2}U^{\dagger}\right)=\nonumber\\&\frac{1.27\times2}{E}\begin{pmatrix}0 & 0 & 0 & 0\\0 & \Delta{m^2} \cos^{2}{\theta} & \Delta{m^2} \sin{\theta} \cos{\theta} & 0\\0 & \Delta{m^2} \sin{\theta} \cos{\theta} & \Delta{m^2} \sin^{2}{\theta} & 0\\0 & 0 & 0 & 0\end{pmatrix}.
\label{Eq:18}
\end{align}
On decomposing this flavour Hamiltonian using Hilbert-Schmidt decomposition mentioned earlier, we get.
\begin{align}
\hat{H}_f=&\frac{1.27\times2}{E}\left[\frac{\Delta{m^2}}{4}(\mathbf{1}\otimes\mathbf{1})\right.\nonumber\\&+\frac{\Delta{m^2}}{4}(\sin^2\theta-\cos^2\theta)(\mathbf{1}\otimes\sigma_z)\nonumber\\
&+\frac{\Delta{m^2}}{2}(\sin\theta\cos\theta)(\sigma_x\otimes\sigma_x)\nonumber\\&+\frac{\Delta{m^2}}{2}(\sin\theta\cos\theta)(\sigma_y\otimes\sigma_y)\nonumber\\
&+\frac{\Delta{m^2}}{4}(\cos^2\theta-\sin^2\theta)(\sigma_z\otimes\mathbf{1})\nonumber\\&\left.-\frac{\Delta{m^2}}{4}(\sigma_z\otimes\sigma_z)\right].
\label{Eq:19}
\end{align}
This construction allows one to simulate the Hamiltonian on a quantum circuit with appropriate two-qubit gates or calculate the unitary time evolution using Lie-Trotter product given in Eq. (\ref{Eq:13}). Additionally, from the decomposition, we get a picture of how these qubits interact. One can immediately see the qubit-qubit exchange interactions along with the free fields. We can remove terms like $\Delta{m^2}(\mathbf{1}\otimes\mathbf{1})/4$ using a unitary transformation, as $(\mathbf{1}\otimes\mathbf{1})$ commutes with every other terms. While designing simulations with ion traps or optical cavities, the coefficients correspond to coupling constants or cavity modes. We will encounter similar construction while introducing matter potentials, and the decomposition becomes more useful for numerical simulation. Since for two flavour approximation in vacuum, we can compute the exact analytical solution for PBP and Eq. (\ref{Eq:19}) serves the purpose of illustration alone. 

Just for convenience we define,
\begin{equation*}
\tilde{\omega}=1.27\left(\frac{\Delta m^2}{E}\right),
\end{equation*}
Then on substituting Eq. (\ref{Eq:18}) to Eq. (\ref{Eq:10}) with, $\mathds{U}(t)\approx\mathds{U}(L)=\exp\left(-i\hat{H}_fL\right)$, we get the unitary time evolution as,
\begin{widetext}
\begin{equation}
\mathds{U}(L) =   \begin{pmatrix}1 & 0 & 0 & 0\\0 & \left( e^{i2\tilde{\omega}L} \sin^{2}{\theta} +\cos^{2}{\theta}\right) e^{-i2\tilde{\omega}L} & \sin{\left(2 \theta \right)}(e^{-i2\tilde{\omega}L}- 1)/2 & 0\\0 & \sin{\left(2 \theta \right)}(e^{-i2\tilde{\omega}L}- 1)/2 &  \left(e^{i2\tilde{\omega}L}\cos^{2}{\theta} + \sin^{2}{\theta}\right) e^{-i2\tilde{\omega}L} & 0\\0 & 0 & 0 & 1\end{pmatrix}.
\label{Eq:20}
\end{equation}
\end{widetext}
Since, $\mathds{U}(L)$ is unitary as it satisfies $\mathds{U}(L)\mathds{U}(L)^{\dagger}=\mathds{1}$, we use Eq. (\ref{Eq:3}) to calculate PBP in the next section. 

To check the validity of our construction, let us consider the initial state $\ket{\nu_{\alpha}(0)}=\ket{10}$. Then the time evolution of state vector is given by,
\begin{equation}
\ket{\nu_{\alpha}(L)}=\mathds{U}(L)\ket{\nu_{\alpha}(0)}
\label{Eq:21}
\end{equation}
On substituting Eq. (\ref{Eq:20}) and taking the inner products $\bra{\nu_{\alpha}}\mathds{U}(L)\ket{\nu_{\alpha}(0)}$ and $\bra{\nu_{\beta}}\mathds{U}(L)\ket{\nu_{\alpha}(0)}$ , we get,
\begin{align}
\mathds{U}_{\alpha\alpha}(L)&=\left(e^{i2\tilde{\omega}L} \sin^{2}{\theta} + \cos^{2}{\theta}\right) e^{-i2\tilde{\omega}L}\label{Eq:22}\\
\mathds{U}_{\alpha\beta}(L)&= \sin{\left(2 \theta \right)}(e^{-i2\tilde{\omega}L}- 1)/2.
\label{Eq:23}
\end{align}
Norm square of the above expressions give the standard oscillation probabilities and $|\mathds{U}_{\alpha\alpha}(L)|^2+|\mathds{U}_{\alpha\beta}(L)|^2=1$. Thus our construction is consistent with the standard PMNS formalism. The above derivation account for the coefficients of Eq. (\ref{Eq:6}). One can repeat the process for three flavour model and construct a flavour Hamiltonian of the form Eq. (\ref{Eq:19}) (See Appendix A).
\subsection{Matter potential in two flavour approximation}
The charged current (CC)  weak interaction can cause coherent forward scattering of electron neutrinos and antineutrinos when they travel through ordinary matter. Works by Mikheyev - Smirnov and independent work by Wolfenstein gives a detailed description of the matter effect in neutrino oscillation. This process is analogous to the refraction of light, and hence an effective mass and mixing angle can be described for the process. 

For two flavour model, we have,
\begin{equation}
V = \text{diag}(0,0,\pm a,0)
\label{Eq:24}
\end{equation}
Where
\begin{equation}
a=2\sqrt{2}G_FN_eE.
\label{Eq:25}
\end{equation}
Here, $G_F$ is the Fermi constant, $N_e$ is the electron number density and $\pm$ corresponds to the situation for neutrinos and antineutrinos respectively. In the original formulation we should have a term $[a/(2E)]^2$. Since $[a/(2E)]<<E$, $[a/(2E)]^2$ is neglected while constructing Eq. (\ref{Eq:14}). Now, similar to Eq. (\ref{Eq:19}) the effective flavour Hamiltonian takes the form,

\begin{align}
\hat{H}_f=&\frac{1.27\times2}{E}\left[\left(\frac{\Delta{m^2\pm a}}{4}\right)(\mathbf{1}\otimes\mathbf{1})\right.\nonumber\\&+\left(\frac{\Delta{m^2}\sin^2\theta-\Delta{m^2}\cos^2\theta\pm a}{4}\right)(\mathbf{1}\otimes\sigma_z)\nonumber\\
&+\frac{\Delta{m^2}}{2}(\sin\theta\cos\theta)(\sigma_x\otimes\sigma_x)\nonumber\\&+\frac{\Delta{m^2}}{2}(\sin\theta\cos\theta)(\sigma_y\otimes\sigma_y)\nonumber\\
&+\left(\frac{\Delta{m^2}\cos^2\theta-\Delta{m^2}\sin^2\theta\mp a}{4}\right)(\sigma_z\otimes\mathbf{1})\nonumber\\&\left.-\left(\frac{\Delta{m^2\mp a}}{4}\right)(\sigma_z\otimes\sigma_z)\right].
\label{Eq:26}
\end{align}
Equation (\ref{Eq:26}) reduces to Eq. (\ref{Eq:19}) when $a=0$. In this case the unitary time evolution of the form Eq. (\ref{Eq:20}) is tedious to handle. Hence we use numerical methods to solve the Schr\"{o}dinger equation.
 
A re-parametrization of $\theta\rightarrow\theta_M$ and $\Delta m^2\rightarrow\Delta m^2_M$ makes $\mathds{U}(\theta,\Delta m^2,L)\rightarrow\mathds{U}(\theta_M,\Delta m^2_M,L_M)$. We can use the standard transformations found in the literature \cite{Giunti:2007ry} given by,
\begin{align}
&\Delta m^2_M=\sqrt{\left[\Delta m^2\cos\left(2\theta\right)-a\right]^2+\left[\Delta m^2\sin\left(2\theta\right)\right]^2}\label{Eq:27}\\
&\tan\left(2\theta_M\right)=\tan\left(2\theta\right)\left(1-\frac{a}{\Delta m^2\cos\left(2\theta\right)}\right)^{-1}
\label{Eq:28}
\end{align}

The reparametrization makes the equation simpler, and the oscillation length gets modified by the vacuum mixing angle. One can either choose to work with $\theta_M,\Delta m^2_M, L_M$ or $\theta,\Delta m^2, L$. Either way, the results will be consistent.  Here, we will use Eq. (\ref{Eq:26}) for our analysis.

\section{Pancharatnam-Berry phase in neutrinos}
Now we are fully equipped to calculate the PBP in neutrinos. Since, we are dealing with pure states under unitary time evolution, we can use Eq. (\ref{Eq:3}) to compute PBP. We will derive each factor in Eq. (\ref{Eq:3}) separately and combine them to find the complete solution. For the time being, let us assume the initial state as,
\begin{equation}
\ket{\psi(0)}=\ket{01}=\begin{pmatrix}
0\\1\\0\\0
\end{pmatrix}.
\label{Eq:29}
\end{equation}
Now, applying the unitary time evolution given by Eq. (\ref{Eq:20}) on the initial state, we get, 
\begin{equation}
\ket{\psi(t)}=\begin{pmatrix}0\\\left(e^{i2\tilde{\omega}t} \sin^{2}{\theta} + \cos^{2}{\theta}\right) e^{-i2\tilde{\omega}t}\\\sin{\left(2 \theta \right)}(e^{-i2\tilde{\omega}t}-1)/2\\0\end{pmatrix}.
\label{Eq:30}
\end{equation}
Here, we haven't replaced $t$ with $L$, as we need to compute the time derivative and integrate it to calculate the dynamical phase factor. From Eq. (\ref{Eq:30}), we get the time derivative as,
\begin{equation}
\frac{\partial}{\partial t}\ket{\psi(t)}=\begin{pmatrix}0\\- i2\tilde{\omega}\cos^{2}{\theta}e^{-i2\tilde{\omega}t}\\- i\tilde{\omega}\sin{\left(2 \theta \right)}e^{-i2\tilde{\omega}t}\\0\end{pmatrix}.
\label{Eq:31}
\end{equation}
Then,
\begin{equation}
\bra{\psi(t)}\partial_t\ket{\psi(t)}=- i\left(\frac{2.54 \Delta{m^2}}{E}\right)\cos^{2}{\theta}.
\label{Eq:32}
\end{equation}
On computing the definite integral given by Eq. (\ref{Eq:1}) for an arbitrary time period $T$ using Eq. (\ref{Eq:32}) we get,
\begin{equation}
\int_{0}^{T}\bra{\psi(t)}\partial_t\ket{\psi(t)}\text{d}t=- i\left(\frac{2.54 \Delta{m^2}T}{E}\right)\cos^{2}{\theta}.
\label{Eq:33}
\end{equation}
This concludes the calculation of the dynamical phase. Further, the total phase is the argument of the following inner product.
\begin{align}
\braket{\psi(0)|\psi(T)}=\left(e^{i2\tilde{\omega}T} \sin^{2}{\theta} + \cos^{2}{\theta}\right)e^{-i2\tilde{\omega}T}.
\label{Eq:34}
\end{align}
Substituting Eq. (\ref{Eq:33}) and Eq. (\ref{Eq:34}) to Eq. (\ref{Eq:3}) we get the PBP as, 
\begin{align}
\Phi=\arg\left[\left(e^{i2\tilde{\omega}T} \sin^{2}{\theta} + \cos^{2}{\theta}\right)e^{-i2\tilde{\omega}T\sin^{2}{\theta}} \right].
	\label{Eq:35}
\end{align}
One can start with a different initial state and compute PBP in the same manner. With,
\begin{equation}
\ket{\psi(0)}=\ket{10}=\begin{pmatrix}
0\\0\\1\\0
\end{pmatrix},
\label{Eq:36}
\end{equation}
we get,
\begin{align}
\Phi=\arg\left[\left(e^{i2\tilde{\omega}T} \sin^{2}{\theta} + \cos^{2}{\theta}\right)e^{-i2\tilde{\omega}T\cos^{2}{\theta}} \right].
\label{Eq:37}
\end{align}
Equation (\ref{Eq:35}) and Eq. (\ref{Eq:37}) gives the general expression for PBP for a time $T$. We can impose the cyclic condition by equating,
\begin{equation}
\frac{2.54\Delta{m^2}T}{E}=2\pi.
\label{Eq:38}
\end{equation}
Since we are already in natural units and we used `$t$' only for calculation purpose, we can write, 
\begin{equation}
T\approx L= \frac{2\pi E}{2.54\Delta{m^2}}.
\label{Eq:39}
\end{equation}
Under Eq. (\ref{Eq:39}), $T$ ($L$) is the time (distance) at which we could recover the initial flavour and $T/2$ ($L/2$) corresponds to the first oscillation maximum. Considering the standard values reported \cite{Esteban2019}, with $\Delta m^2=2.525\times10^{-3}$ eV$^2$ for $E=2$ GeV, we have the first oscillation maximum at $L=979.7$ km.

Equation (\ref{Eq:35}) and Eq. (\ref{Eq:37}), with modulo $2\pi$, reduce to the results obtained by Blasone \textit{et al.} \cite{BLASONE1999262}, under cyclic condition given by Eq. (\ref{Eq:39}).  Here, the PBP reduces to,
\begin{equation}
\Phi=\left\{\begin{matrix}
2\pi\sin^2\theta&\text{for flavour}\,\ket{01}\\
2\pi\cos^2\theta&\text{for flavour}\,\ket{10}
\end{matrix}\right..
\label{Eq:40}
\end{equation}
Blasone \textit{et al.} \cite{BLASONE1999262} concluded that the PBP is independent of neutrino energies and masses. This observation is a special case of our results. 

Wang \textit{et al.} \cite{PhysRevD.63.053003}, obtained the non-cyclic PBP for neutrinos by considering a new state, given by, $\ket{\tilde{\nu}(t)}=\exp\left[i\int_{0}^{t}\braket{E}(t')dt'\right]\ket{\nu(t)}$, and imposing the condition, $\braket{\nu(0)|\tilde{\nu}(t)}\equiv r\exp(i\beta)$. Dixit \textit{et al.} \cite{Dixit_2018} used the exact formulation by Wang \textit{et al.} for the calculation of PBP, but switched to the formalism illustrated by Ohlsson \textit{et al.} \cite{doi:10.1063/1.533270} to include matter effect. In our treatment, one can incorporate the matter potential by transforming Eq. (\ref{Eq:20}) using Eq. (\ref{Eq:27}) and Eq. (\ref{Eq:28}) or by invoking the numerical methods illustrated. 

Since our analysis treats the total and dynamical phase separately, additional features like nodal points can be recognized in the parameter space. Furthermore, it is interesting to note that Eq. (\ref{Eq:35}) resembles the PBP obtained by Jisha \textit{et al.} \cite{PhysRevA.95.023823}, where they proved that the effective photonic potential due to the spin-orbit interaction originates from the periodic modulation of the Pancharatnam-Berry phase. Recently, Enomoto \textit{et al.} \cite{PhysRevD.99.036005} used a similar framework to explore the features of spontaneous baryogenesis. Thus, our research points out the scope of additional hidden features of PBP in neutrino sectors, which needs further investigations. In the next section, we explore the features shown by our results. For all calculations, we use Eq. (\ref{Eq:29}) as our initial state unless otherwise specified.

\subsection{PBP \& Mixing angle}
Since we have our exact equations for PBP, let us analyse the features of each parameters. From Eq. (\ref{Eq:22}), the oscillation probability in natural units takes the from,
\begin{equation}
\mathds{P}_{\alpha\rightarrow\beta}=|\mathds{U}_{\alpha\beta}(L)|^2=\sin^2(2\theta)\sin^2\left(\frac{1.27\Delta m^2L}{E}\right).
\label{Eq:41}
\end{equation}
Considering the variation along $L$, `$\sin^2(2\theta)$' corresponds to the oscillation amplitude and `$\sin^{2}\left(1.27\Delta m^2L/E\right)$' corresponds to the phase. Then we get the oscillation maximums for,
\begin{equation}
\sin\left(\frac{1.27\Delta m^2L}{E}\right)=1.
\label{Eq:42}
\end{equation}
The oscillation extrema corresponds to,
\begin{equation}
\frac{1.27\Delta m^2L}{E}=n\left(\frac{\pi}{2}\right)\implies L=\frac{n\pi E}{2.54\Delta m^2}
\label{Eq:43}
\end{equation}
For even values of $n$, we get oscillation minimums and for odd values we get maximums. With $E=2$ GeV and $\Delta m^2=2.525\times10^{-3}$ eV$^2$ \cite{Esteban2019}, we have, for $n=1$, $L\approx979.7$ km.

From Eq. (\ref{Eq:34}) the total phase is given by,
\begin{align}
\arg&\left(\braket{\psi(0)|\psi(L)}\right)\nonumber\\&=\arg\left[\left(e^{i2\tilde{\omega}L} \sin^{2}{\theta} + \cos^{2}{\theta}\right)e^{-i2\tilde{\omega}L}\right]
\label{Eq:44}
\end{align}
Considering the $1^{\text{st}}$oscillation maximum based on Eq. (\ref{Eq:43}), the total phase reduces to, 
\begin{equation}
\arg{\left(- \cos{\left(2 \theta \right)} \right)}=\left\{\begin{matrix}
0&\text{for} \,\cos(2\theta)<0,\\
\text{undefined}&\text{for} \,\cos(2\theta)=0,\\
\pi&\text{for} \,\cos(2\theta)>0.
\end{matrix}\right.
\label{Eq:45}
\end{equation}
When $\braket{\psi(0)|\psi(L)}$ changes by a sign, there is a sudden change of phase by a value $\pi$. This change in the total phase also gets reflected in PBP. The points at which $\braket{\psi(0)|\psi(L)}=0$ are the nodal points. For illustration clarity, we will plot the phases defined in the range $[-\pi,\pi]$. One can convert the phase defined in the region $[-\pi,\pi]$ to $[0,2\pi]$ by taking the modulo $2\pi$. The contrast between the definition of the range appears only in the values. We shall also use the range $[0,2\pi]$ when needed. 

\begin{figure}
	\subfloat[\label{fig:1a}$L=900.0$ km]{\includegraphics[width=0.99\columnwidth]{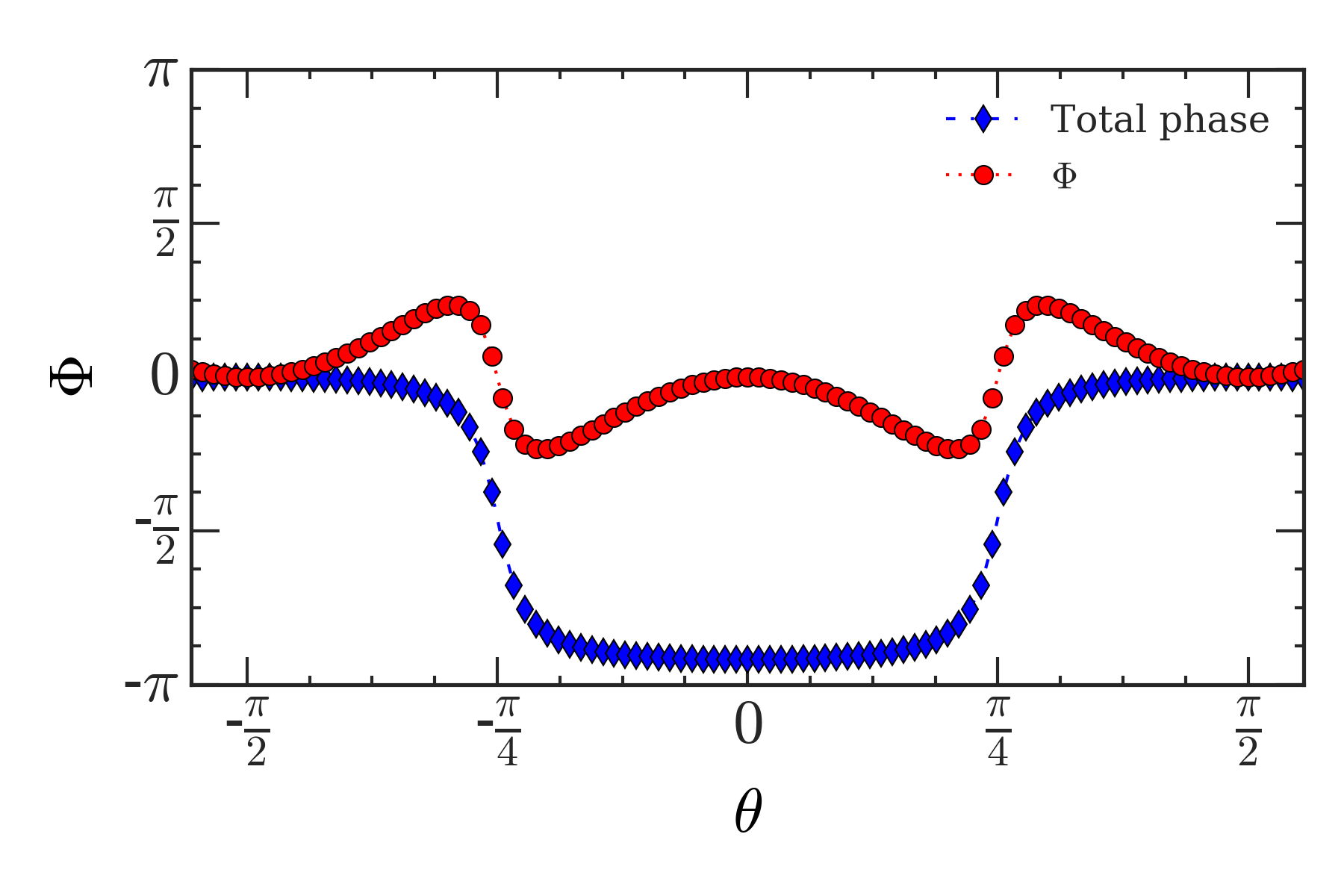}}\\
	\subfloat[\label{fig:1b}$L=979.7$ km]{\includegraphics[width=0.99\columnwidth]{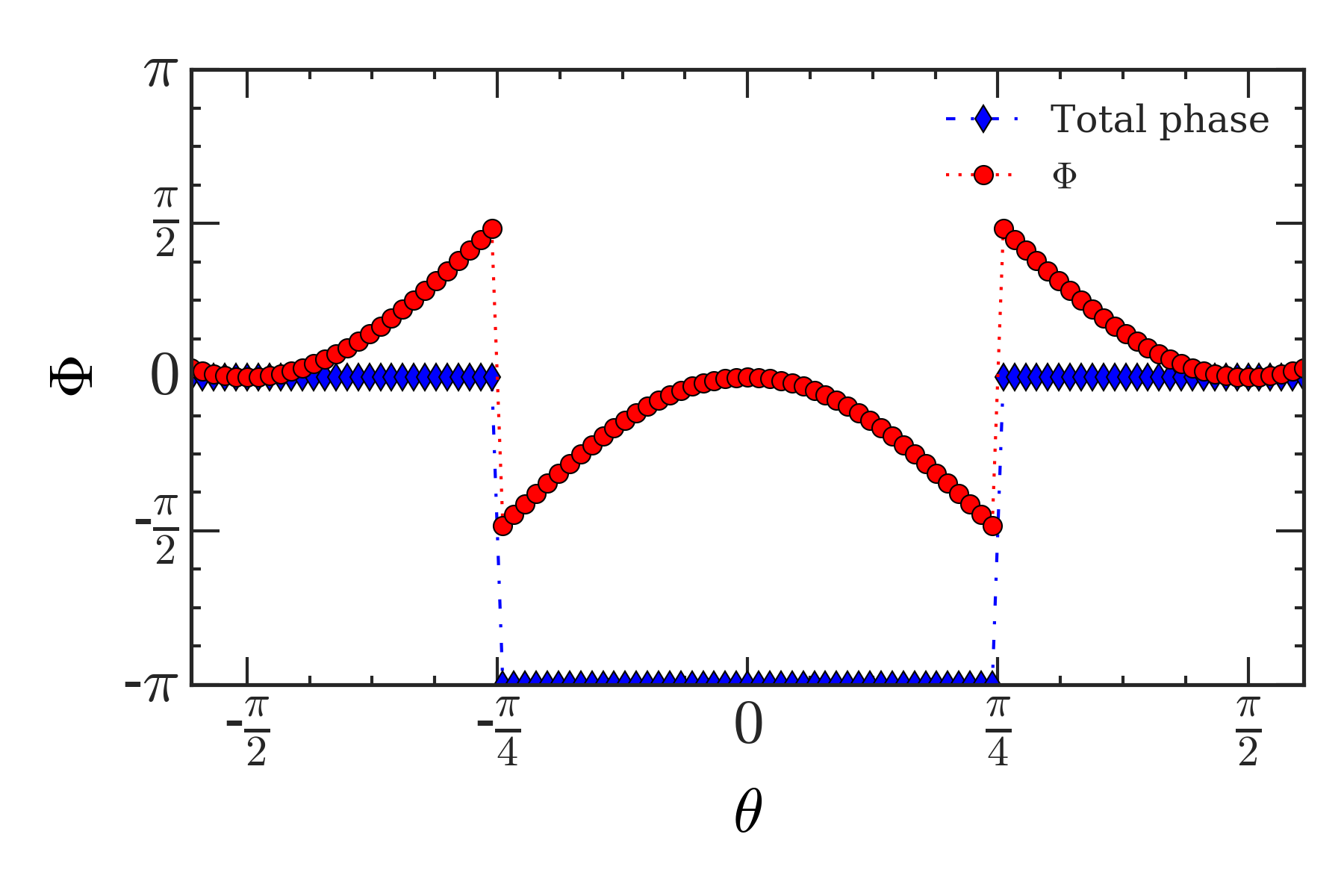}}\\
	\subfloat[\label{fig:1c}$L=1959.4$ km]{\includegraphics[width=0.99\columnwidth]{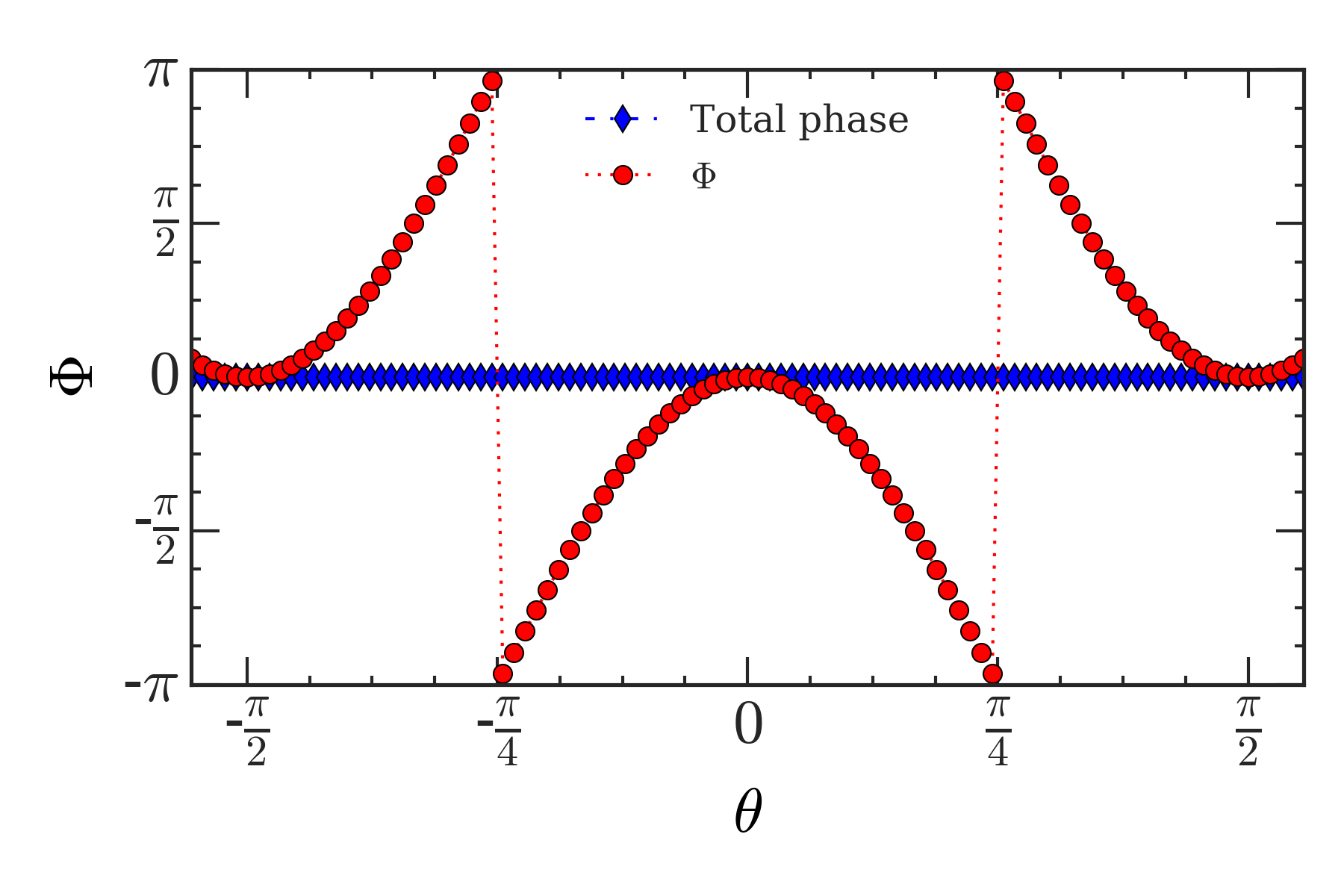}}
	\caption{(Colour online) PBP and total phase plotted against $\theta$, with $E=2$ GeV and $\Delta m^2=2.525\times10^{-3}$ eV$^2$.}
	\label{fig:1}
\end{figure}

Figure (\ref{fig:1}) illustrates the behaviour of PBP ($\Phi$) and total phase for different baselines. In Fig. (\ref{fig:1a}), we have $L=900$ km, which is less than the $1^{\text{st}}$ oscillation maximum, and we still have a non zero PBP. At $1^{\text{st}}$ oscillation maximum, Fig. (\ref{fig:1b}), we can see a sudden jump of $\pi$ for the total phase as described in Eq. (\ref{Eq:45}). This jump occurs at $\theta=\pi/4$, which is the same as the angle in bi-maximal mixing. Such points are called the nodal points. For bi-maximal mixing, the PBP acquires a value of $\pm\pi$ ($\pm\pi$  modulo $2\pi$ corresponds to $\pi$) after a cyclic evolution. The diagram tells the correlation between PBP and mixing angles, which was clear from Eq. (\ref{Eq:35}) and Eq. (\ref{Eq:37}).

It is interesting to note that, for $1^{\text{st}}$ oscillation maximum, the total phase can only take values either zero or $\pi$ (on taking modulo 2$\pi$). This observation was the main result by Mehta \cite{PhysRevD.79.096013}. This behaviour is due to the presence of nodal points and the expression of total phase in Eq. (\ref{Eq:45}). The physical significance of nodal points will be made clear while we discuss MSW resonance. Further, when we consider a complete cycle, given by the cyclic condition in Eq. (\ref{Eq:38}), the state returns to its initial state and the total phase vanishes (see Fig. (\ref{fig:1c})). This cyclic path corresponds to a closed circuit in ray space, and the magnitude of PBP becomes equal to the magnitude of the dynamical phase. 
\begin{figure}[h]
	\includegraphics[width=1\columnwidth]{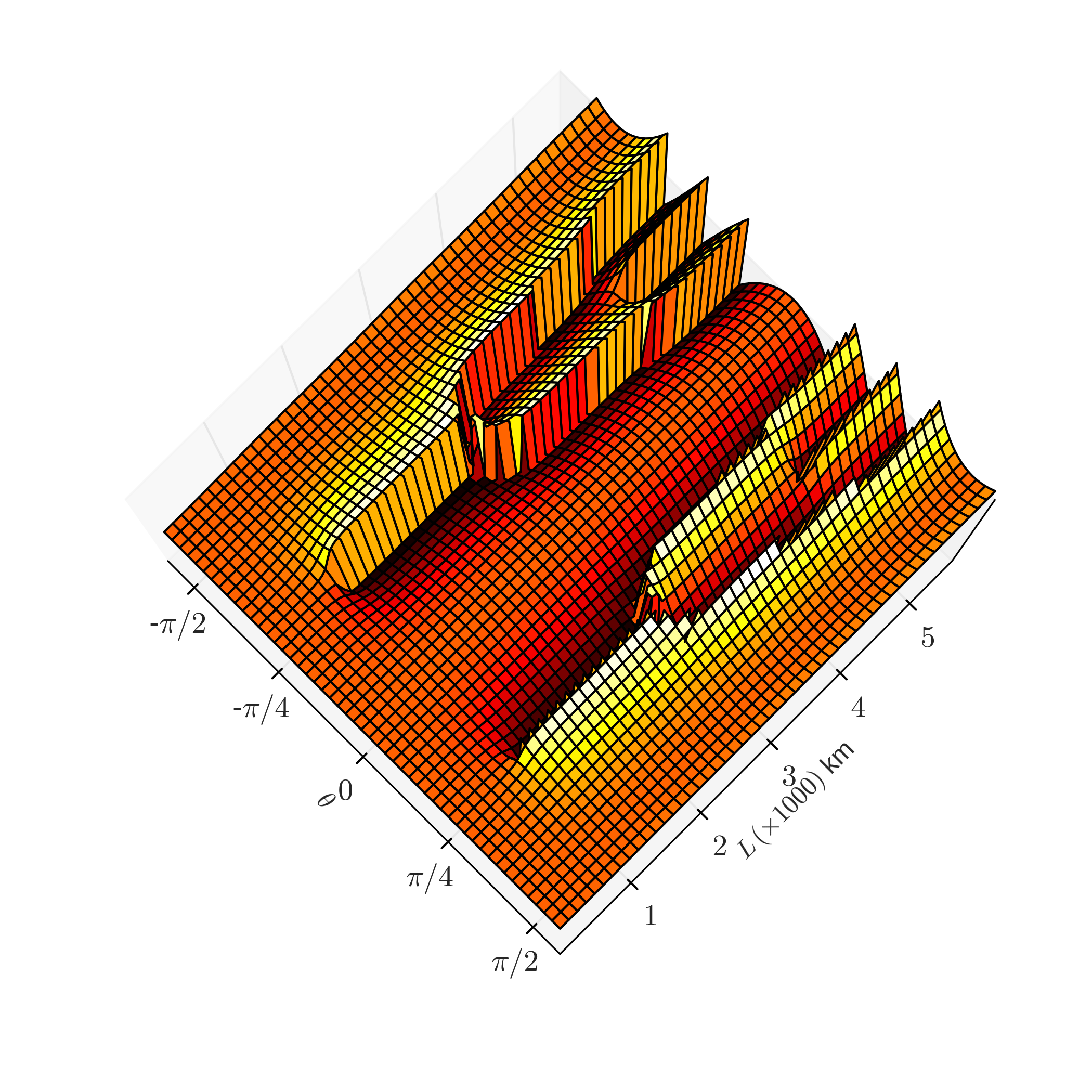}
	\caption{(Colour online) PBP for various values of $\theta$ and $L$ with $E=2$ GeV and $\Delta m^2=+2.525\times10^{-3}$ eV$^2$. $z$ axis corresponds to PBP and ranges from $[-\pi,\pi]$}
	\label{fig:2}
\end{figure}
\begin{figure}
	\includegraphics[width=0.99\columnwidth]{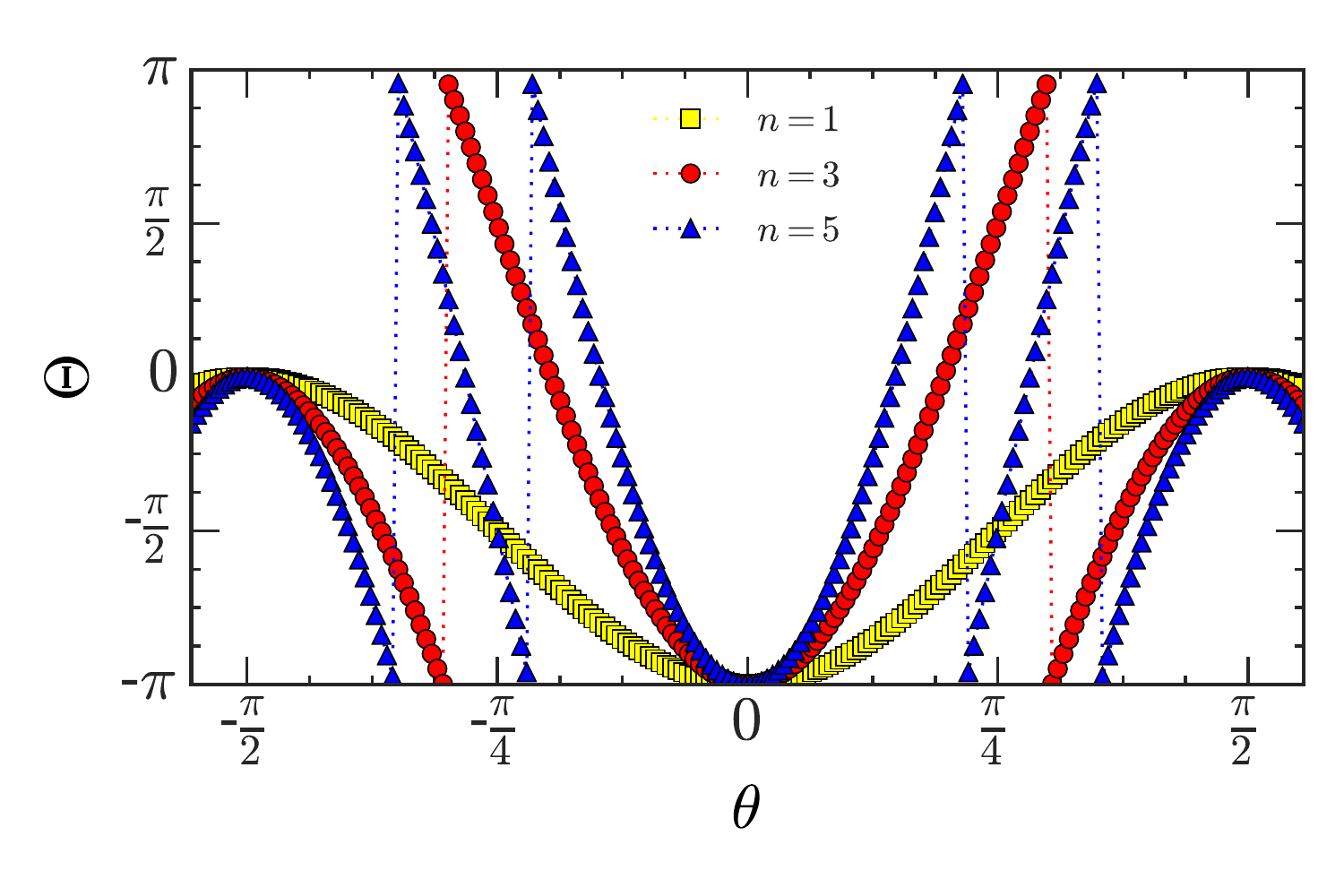}
	\caption{(Colour online) Dynamical phase for $n^{\text{th}}$ oscillation extremum from Eq. (\ref{Eq:46}).}
	\label{fig:3}
\end{figure}
Additionally, it is interesting to point out bifurcation points for each oscillation maxima in Fig. (\ref{fig:2}). These bifurcations are not due to additional nodal points, but due to the behaviour of the dynamical phase. Dynamical phase for $n^{\text{th}}$ oscillation extremum is,
\begin{equation}
\Theta=\arg \left[\exp\left\{-i\frac{1}{4}  \pi  \left(e^{-2 i \theta }+e^{2 i \theta }+2\right) n\right\}\right]
\label{Eq:46}
\end{equation}
In Fig. (\ref{fig:3}), we can see the jump of $2\pi$ at different values of $\theta$, by different number of times for different values of $n$. In all these cases, there are no nodal points, which corresponds to a shift of $\pi$. 
\begin{figure}[h]
	\includegraphics[width=1\columnwidth]{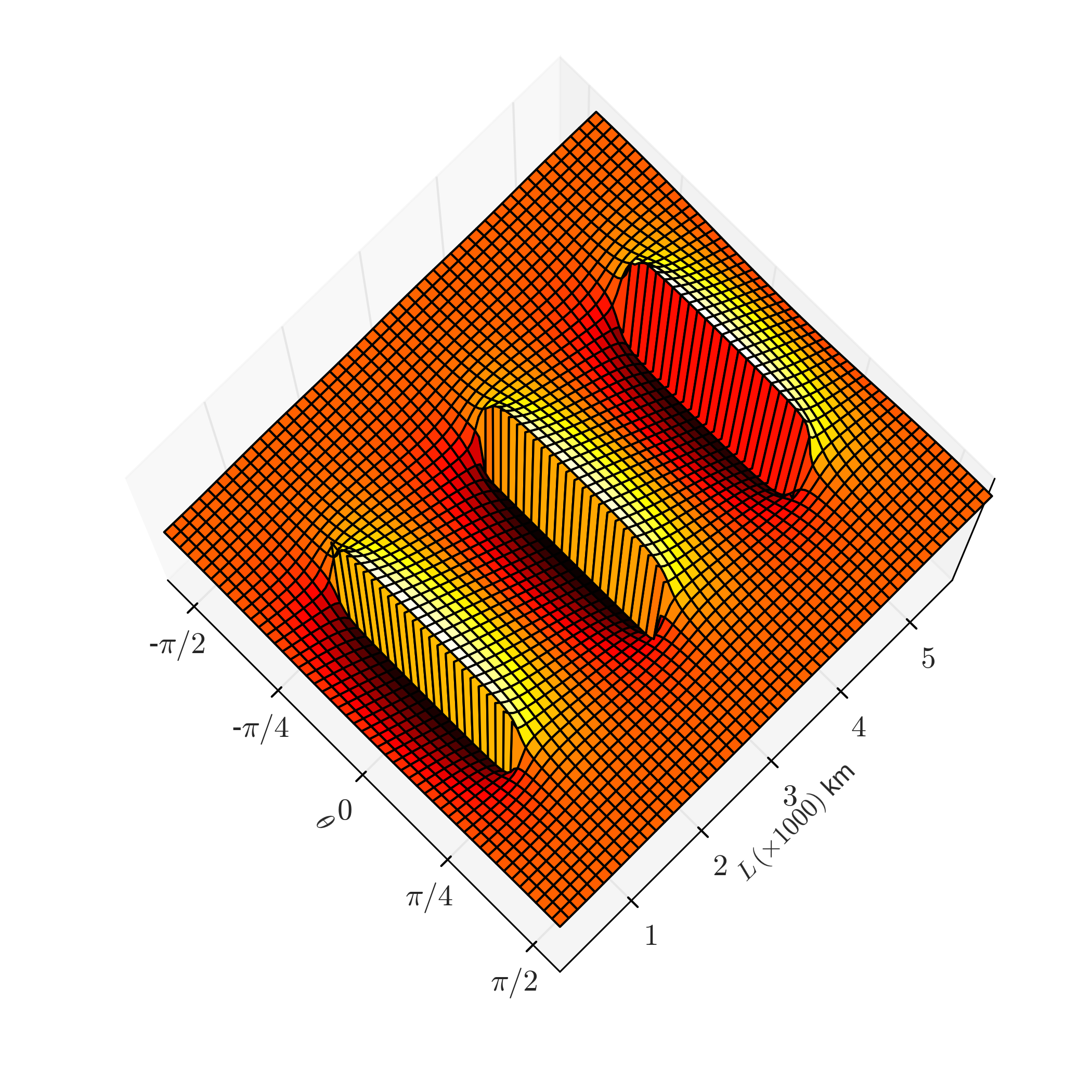}
	\caption{(Colour online) Total phase factor for various values of $\theta$ and $L$ with $E=2$ GeV and $\Delta m^2=+2.525\times10^{-3}$ eV$^2$. $z$ axis corresponds to total phase and ranges from $[-\pi,\pi]$.}
	\label{fig:4}
\end{figure}
In fact, the nodal points arise due to the total phase and it corresponds to $\theta\in\{-\pi/4,\pi/4\}$ in vacuum (see Fig. (\ref{fig:4})). For even values of $n$, one of the $2\pi$ jumps coincides with the nodal points, and for odd values, we get the different bifurcations due to $2\pi$ jumps. For non integer values of $n$, the curve in Fig. (\ref{fig:3}) assumes intermediate values. 

\subsection{PBP \& Mass squared difference}
\begin{figure}[ht]
	\subfloat{\includegraphics[width=0.99\columnwidth]{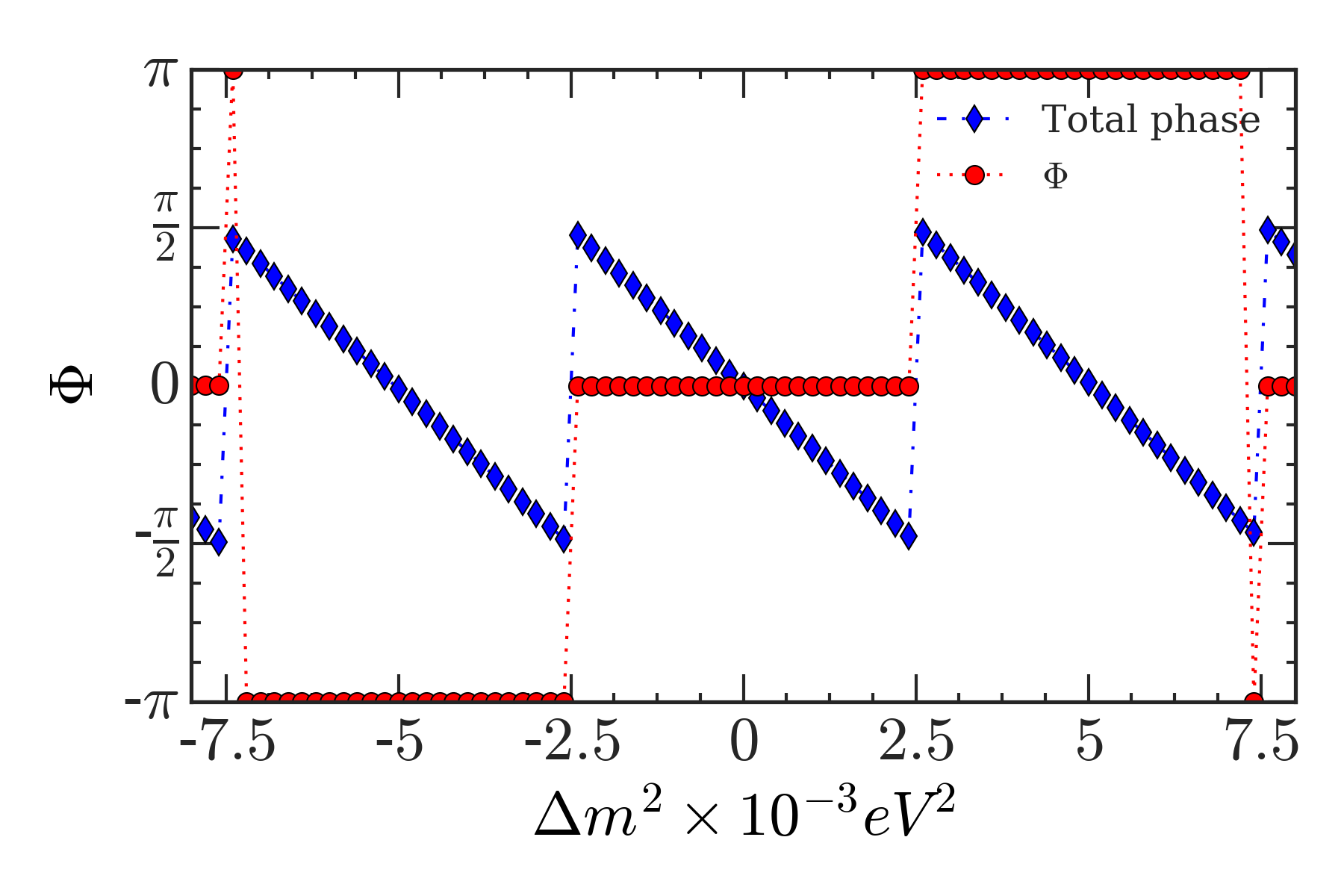}}
	\caption{(Colour online) PBP and the total phase plotted against $\Delta m^2$, with $E=2$ GeV,  $L=979.7$ km and $\theta=45^o$.}
	\label{fig:5}
\end{figure}
\begin{figure}[ht]
	\includegraphics[width=1\columnwidth]{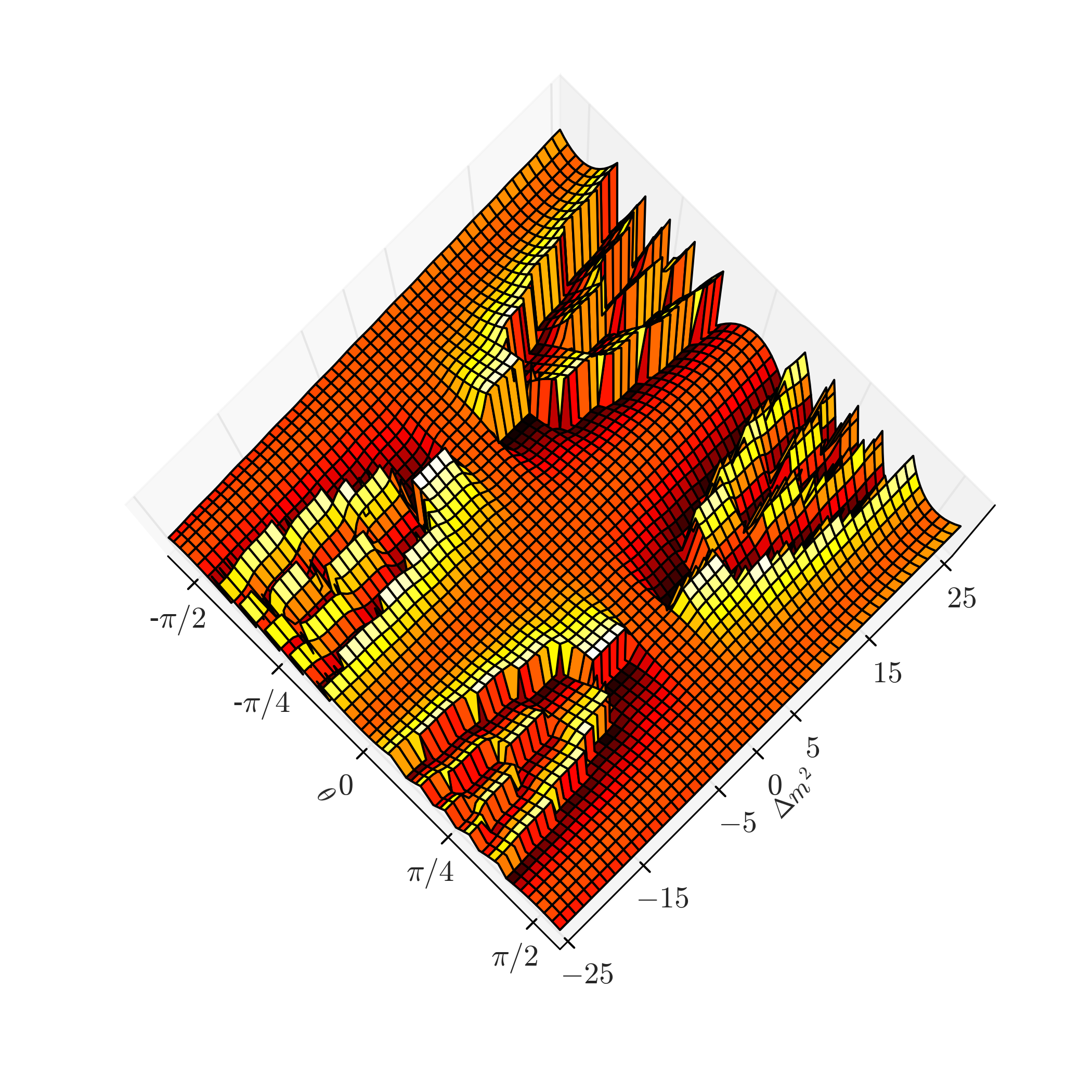}
	\caption{(Colour online) PBP for various values of $\theta$ and $\Delta m^2$, with $E=2$ GeV and  $L=979.7$ km. $z$ axis corresponds to PBP and ranges from $[-\pi,\pi]$.}
	\label{fig:6}
\end{figure}
So far, we have kept the value of $\Delta m^2=+2.525\times10^{-3}$ eV$^2$, which corresponds to the normal mass hierarchy in the literature. It is natural to ask what happens when an inverted hierarchy is chosen. In Fig. (\ref{fig:5}), we can see the nodal jumps of $\pi$ at $2.525\times10^{-3}$ eV$^2$ for $L=979.7$ km and $\theta=45^0$. Additionally, jumps of $2\pi$ exist for the conditions manifested by the dynamical part. Figure (\ref{fig:6}) shows the behaviour of PBP for various values of $\theta$ and $\Delta m^2$ for a given oscillation length. The most important feature to notice is the difference in PBP for different mass hierarchies. Sensitive measurement of PBP can identify the mass hierarchy.
\begin{figure}
	\includegraphics[width=0.99\columnwidth]{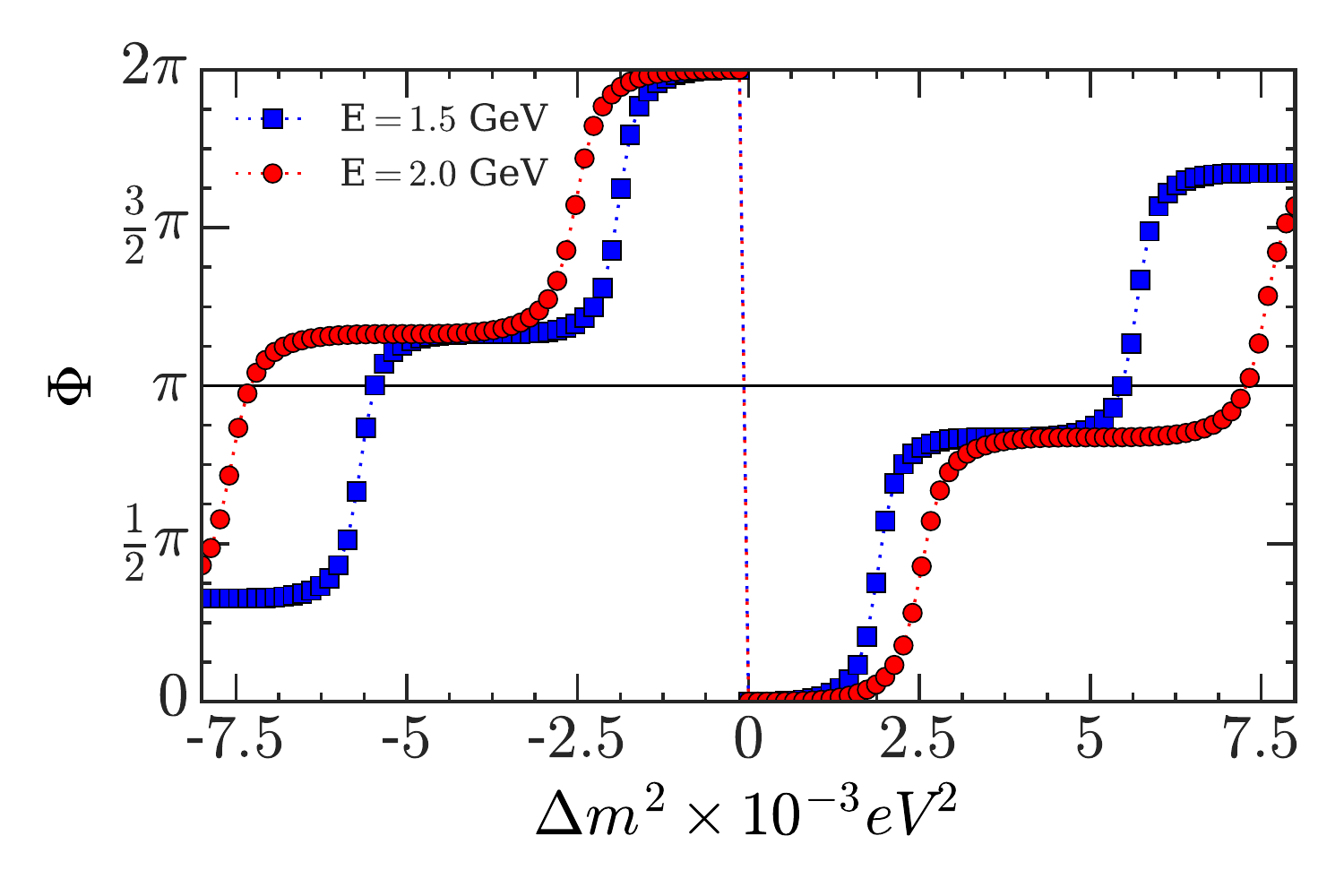}
	\caption{(Colour online) PBP for various values of $E$ and $\Delta m^2$, with  $L=979.7$ km and $\theta=49.7^o$.}
	\label{fig:7}
\end{figure}

In Fig. (\ref{fig:7}), we have redefined PBP between $[0,2\pi]$, and we can see the difference between normal and inverted mass hierarchies. Although the numerical value of the mass squared difference ends up in a PBP plateau, it is sensitive to the neutrino's total energy ($E$). At bi-maximal mixing ($\theta=45^0$) these difference vanishes and the PBP plateaus converge to $\pi$. Since we have strong evidence against bi-maximal mixing, such degeneracy can be excluded. Thus, PBP plateaus are sensitive to the total energy ($E$). It is possible to determine mass hierarchy by measuring PBP and putting numerical bounds on the value of $\Delta m^2$.

\subsection{PBP \& Matter potential}
In our description matter effects can be included by the transformation $\{\theta,\Delta m^2\}\rightarrow\{\theta_M,\Delta m^2_M\}$ given  by Eq. (\ref{Eq:27}) and Eq. (\ref{Eq:28}). Since, $L_M$ is a function of $\Delta m^2_M$ as, $L_M=n\pi E/(2.54\Delta m^2_M)$ and $\Delta m^2_M$ is a function of $\theta$ for non zero $a$, the vacuum mixing angle affects the oscillation length. Therefore it is convenient to use the Eq. (\ref{Eq:26}) and solve the system numerically, by which we can express the results in the vacuum oscillation parameters and examine the features of nodal points. 
\begin{figure}[ht]
	\includegraphics[width=0.99\columnwidth]{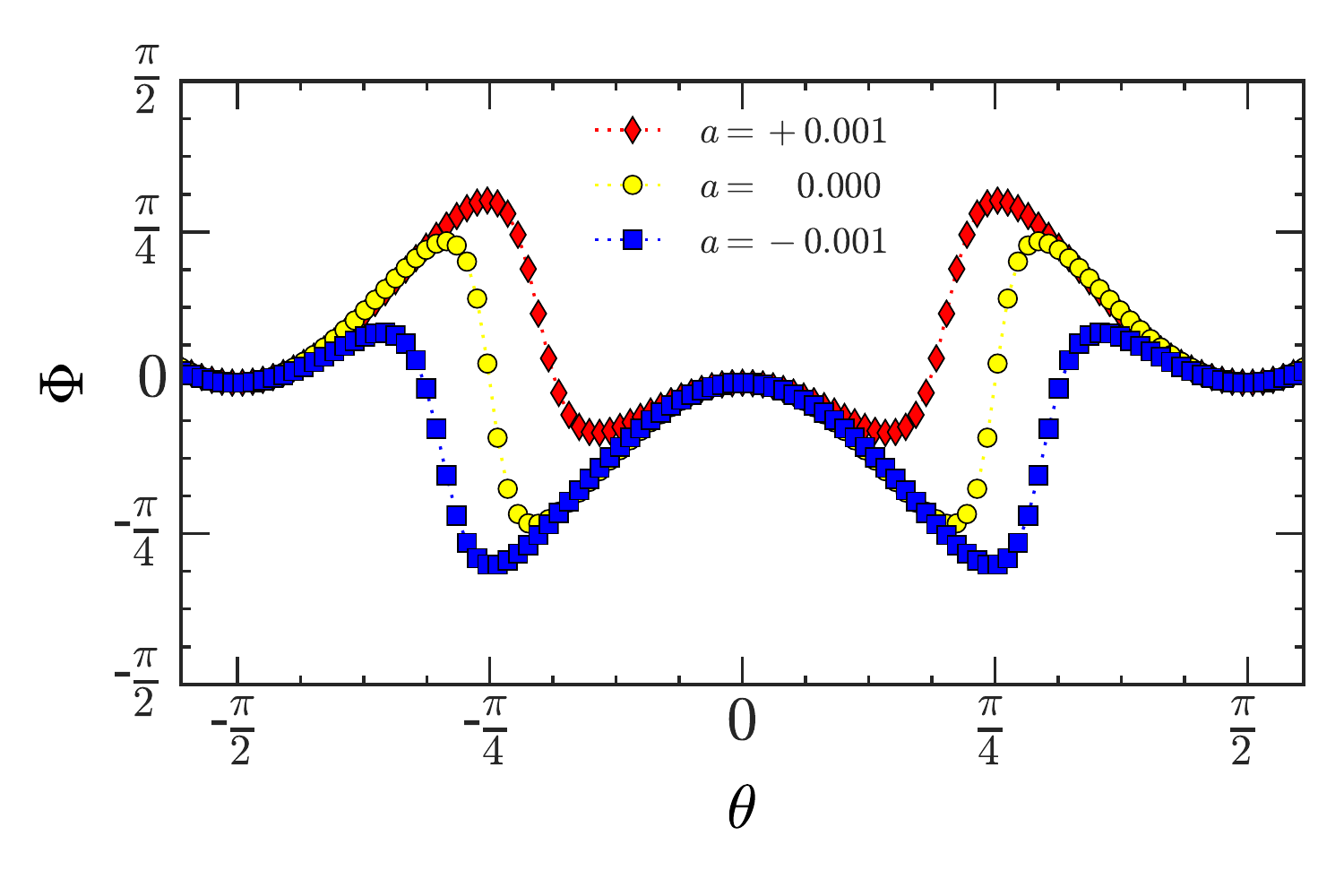}\\
	\caption{(Colour online) PBP plotted against $\theta$ for different values of $a$ (eV$^2$), with $\Delta m^2=+2.525\times10^{-3}$ eV$^2$ and $L=900$ km.}
	\label{fig:8}
\end{figure}

Figure (\ref{fig:8}) illustrates the effects of matter potential on neutrinos ($a=+$) and antineutrinos ($a=-$). It is evident that, PBP is sensitive to neutrinos and antineutrinos. With matter potential, nodal points shift from $\pi/4$ to values less than or greater than $\pi/4$ depending on the sign of $a$. It is interesting to note that, for $a>0$, the nodal points are at $|\theta|<\pi/4$ and for $a<0$, the nodal point are at $|\theta|>\pi/4$. These inequalities account for the MSW effect. For two flavour approximation, in the absence of matter, the bi-maximal mixing is the resonance oscillation with probability amplitude ($\mathds{P}$) reaching its maximum value. In the presence of matter, for neutrinos, the resonance can appear only for $\theta<\pi/4$ and in the case of antineutrinos, with a reversed potential, this resonance can occur if $\theta>\pi/4$. From our analysis, it is clear that PBP also recognizes such resonance points. In other words, the nodal points appear at the resonance for which the value of $\cos(2\theta_M)$ in total phase takes a sign change as shown in Eq. (\ref{Eq:45}). 

\begin{figure}[h]
	\subfloat[$L=1066.93$ km]{\includegraphics[width=1\columnwidth]{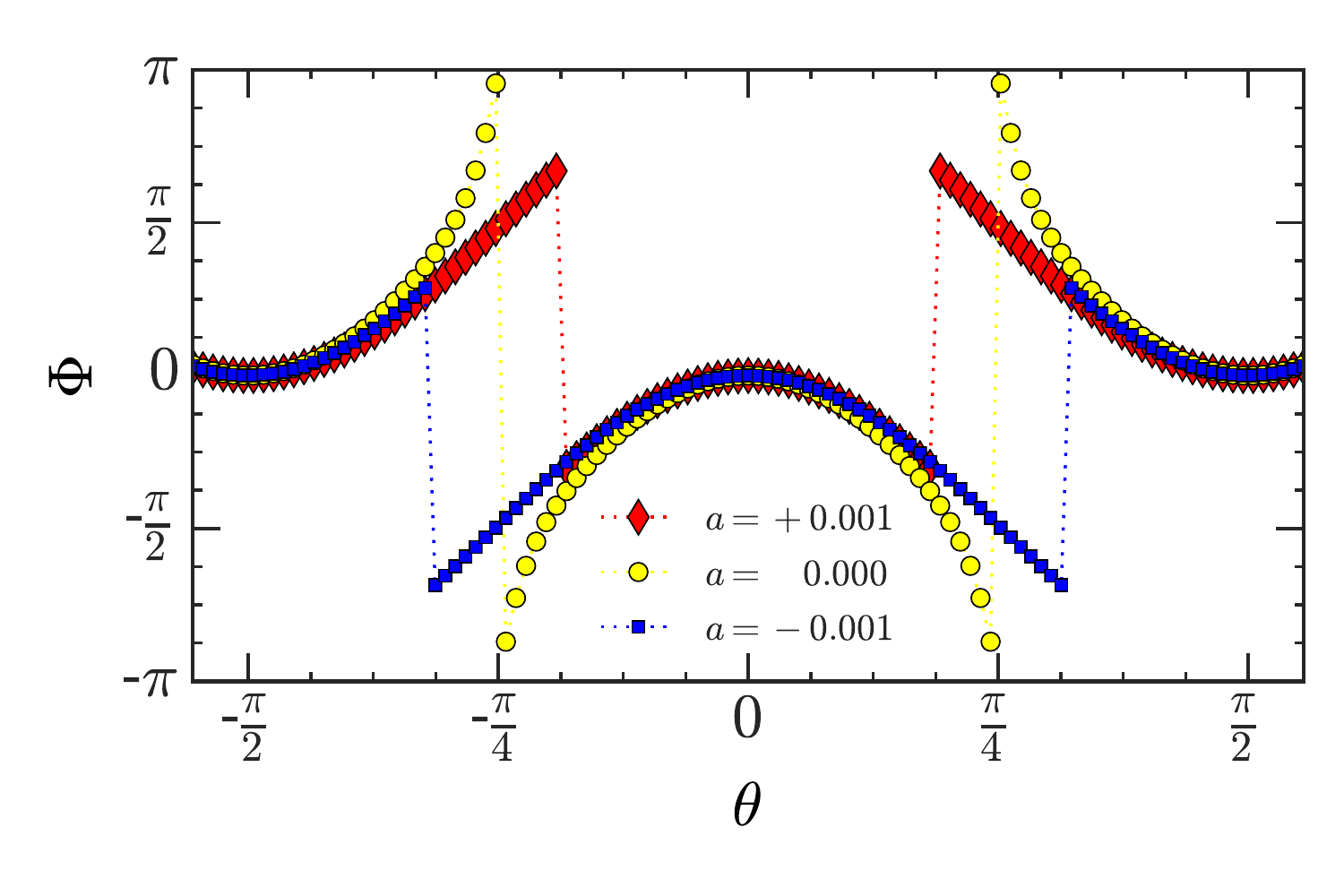}}\\
	\subfloat[$L=1959.4$ km]{\includegraphics[width=1\columnwidth]{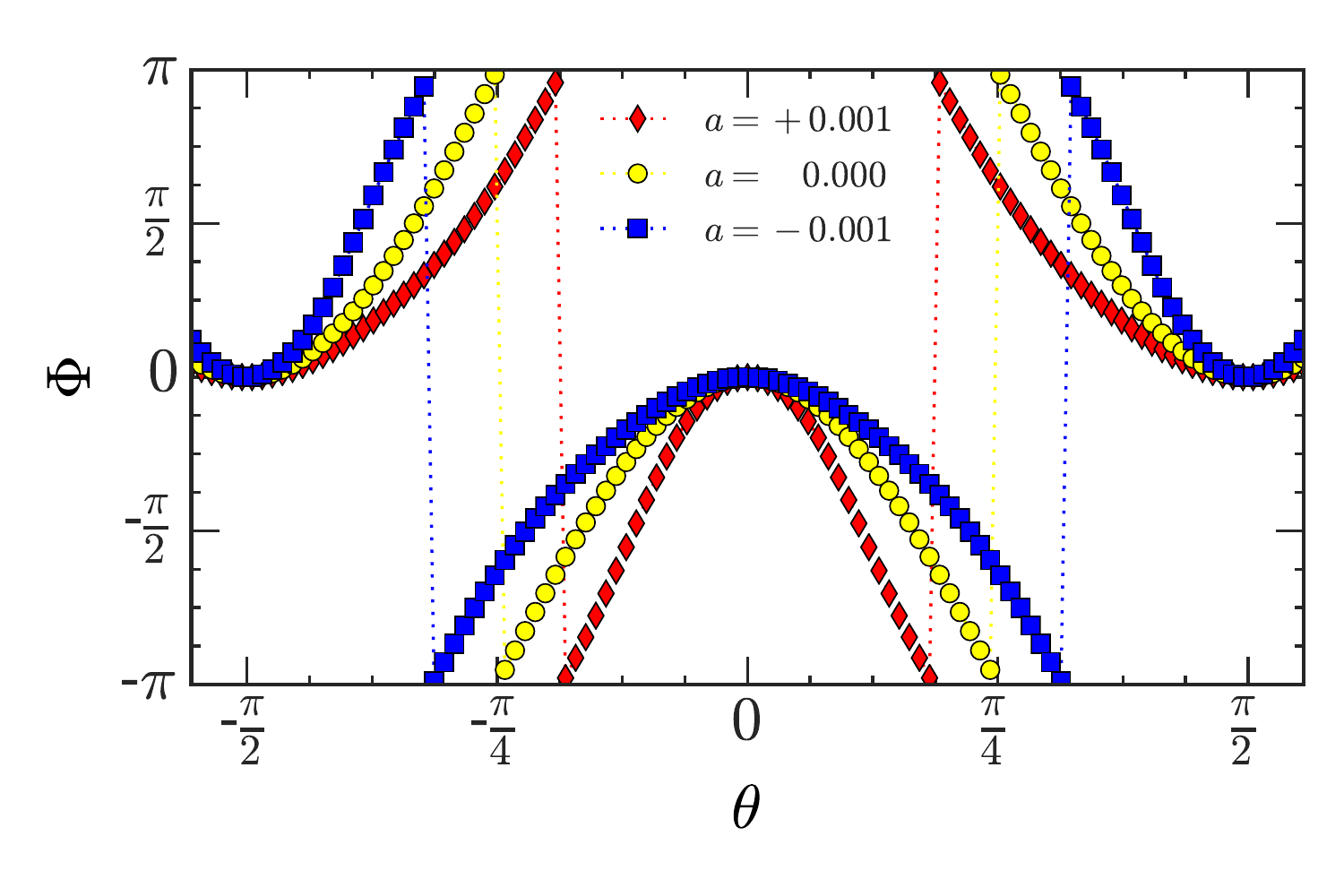}}
	\caption{(Colour online) PBP plotted against $\theta$ for different values of $a$ (eV$^2$) and $L$, with $\Delta m^2=+2.525\times10^{-3}$ eV$^2$.}
	\label{fig:9}
\end{figure}

At resonance, we have $\Delta m^2\cos(2\theta)=a$. This implies, $\theta=\arccos(a/\Delta m^2)/2$. For $a=+0.001$ eV$^2$, we have $\theta=33.34^o$ and for $a=-0.001$ eV$^2$, we have $\theta=56.67^o$. Whenever we have $\Delta m^2\cos(2\theta)=a$, $\theta_M=\pm45^o$ indicating a nodal point in the transformed system. Furthermore, $\Delta m^2_M=\pm\Delta m^2\sin(2\theta)=\pm2.318\times10^{-3}$ eV$^2$. The modified oscillation maximum will be at $L_M=\pi E/(2.54\Delta m^2_M)=1066.9$ km. In Fig. (\ref{fig:9}), we can see the shift in nodal points concerning different matter potentials. And is consistent with the definition of MSW resonance for neutrinos and antineutrinos. This gives physical importance to nodal points of PBP in the context of neutrinos. This research point can be extended to varying matter densities and investigate the effects of terms proportional to the derivative mixing angles.

\subsection{PBP \& The Dirac phase}
From the construction of PMNS formalism, we need at least three flavour model to include the Dirac $CP$ phase ($\delta_{CP}$). This fact is evident from the Jarlskog invariant ($J$), which includes sines and cosines of all mixing angles and the sine of the Dirac $CP$ phase. Thus, making one angle zero makes the invariant vanish, and we cannot distinguish between neutrinos and antineutrinos in oscillation experiments. In the standard PMNS formalism, $\delta_{CP}$ can be measured from the difference in oscillation probabilities of neutrinos and antineutrinos.

The difference between neutrinos and antineutrinos is evident from two parts of the flavour Hamiltonian given by Eq. (\ref{Eq:14}), namely the PMNS matrix ($U$) and the matter potential ($\hat{V}$). For neutrinos, we have $\hat{H}_f$ given by Eq. (\ref{Eq:14}) itself. To construct the same for antineutrinos, we need to complex conjugate $\hat{H}_f$. Since, $\hat{H}_{m^2}$ and $\hat{V}$ are real Hermitian matrices, the only change by taking the complex conjugate comes in $U$, where the factor $e^{i\delta_{CP}}\rightarrow e^{-i\delta_{CP}}$, which is equivalent to $e^{i\delta_{CP}}\rightarrow e^{i(-\delta_{CP})}$. To account for the difference between neutrinos and antineutrinos, we need to analyze $\delta_{CP}$ and $-\delta_{CP}$. Secondly, neutrinos and antineutrinos feel different matter potentials, which only differs by a sign. Here, for ordinary matter, $a$ is $+ve$ for neutrinos and $a$ is $-ve$ for antineutrinos. The effects of $\pm$ were already discussed for two flavour model, and the physical significance of nodal points was also addressed in the last section.

Following the same construction used for two flavour approximation, we can now map our three-level system into three two-level systems. Similar to the previous case, we restrict ourselves to single-mode excitations of the flavour vacuum. In this construction, we will have an $8\times8$ matrices representing our PMNS rotation and the flavour Hamiltonian. Repeating the point, we emphasise the fact that, such an architecture allows one to explore correlation among flavour modes. Here, we have,
\begin{equation}
U=\begin{pmatrix}
1&0&0&0&0&0&0&0\\
0&U_{\tau3}&U_{\tau2}&0&U_{\tau1}&0&0&0\\
0&U_{\mu3}&U_{\mu2}&0&U_{\mu1}&0&0&0\\
0&0&0&1&0&0&0&0\\
0&U_{e3}&U_{e2}&0&U_{e1}&0&0&0\\
0&0&0&0&0&1&0&0\\
0&0&0&0&0&0&1&0\\
0&0&0&0&0&0&0&1\\
\end{pmatrix}.
\label{Eq:47}
\end{equation}
One can immediately see the resemblance with the standard PMNS mixing matrix found in the literature, where (with $\delta_{CP}=\delta$),
\begin{align}
U_{\tau3}=&  \cos{\theta_{13}} \cos{\theta_{23}}\nonumber\\\nonumber
U_{\tau2}=&  - e^{ i \delta} \sin{\theta_{12}} \sin{\theta_{13}} \cos{\theta_{23}} - \sin{\theta_{23}} \cos{\theta_{12}}\nonumber\\
U_{\tau1}=&  - e^{ i \delta} \sin{\theta_{13}} \cos{\theta_{12}} \cos{\theta_{23}} + \sin{\theta_{12}} \sin{\theta_{23}}\nonumber\\
U_{\mu3}=&  \sin{\theta_{23}} \cos{\theta_{13}}\nonumber\\
U_{\mu2}=&  - e^{i \delta} \sin{\theta_{12}} \sin{\theta_{13}} \sin{\theta_{23}} + \cos{\theta_{12}} \cos{\theta_{23}}\nonumber\\
U_{\mu1}=&  - e^{ i \delta} \sin{\theta_{13}} \sin{\theta_{23}} \cos{\theta_{12}} - \sin{\theta_{12}} \cos{\theta_{23}}\nonumber\\
U_{e3}=&  e^{- i \delta} \sin{\theta_{13}}\nonumber\\
U_{e2}=&  \sin{\theta_{12}} \cos{\theta_{13}}\nonumber\\
U_{e1}=&  \cos{\theta_{12}} \cos{\theta_{13}}.
\label{Eq:48}
\end{align}
Now, with the definition followed from Eq. (\ref{Eq:14}), with $\hat{V}=0$ we have,
\begin{align}
\hat{H}_f=\frac{1.27\times2}{E}\begin{pmatrix}
0&0&0&0&0&0&0&0\\
0&H_{\tau\tau}&H_{\tau\mu}&0&H_{\tau e}&0&0&0\\
0&H_{\mu\tau}&H_{\mu\mu}&0&H_{\mu e}&0&0&0\\
0&0&0&0&0&0&0&0\\
0&H_{e\tau}&H_{e\mu}&0&H_{e e}&0&0&0\\
0&0&0&0&0&0&0&0\\
0&0&0&0&0&0&0&0\\
0&0&0&0&0&0&0&0\\
\end{pmatrix},
\label{Eq:49}
\end{align}
where, for any given lepton flavour combination, say $\alpha$ and $\beta$, with $\alpha\in\{\tau, \mu, e\}$ and $\beta\in\{\tau, \mu, e\}$, the matrix elements has the general form given by,
\begin{align}
H_{\alpha\beta}&=U_{\alpha3}\Delta m^2_{31}U^*_{\beta3}+U_{\alpha2}\Delta m^2_{21}U^*_{\beta2}.  
\label{Eq:50}
\end{align}
The exact formulas for each matrix elements in Eq. (\ref{Eq:49}) are given in Appendix A. With a non zero matter potential, the general formula becomes,
\begin{align}
H_{\alpha\beta}&=U_{\alpha3}\Delta m^2_{31}U^*_{\beta3}+U_{\alpha2}\Delta m^2_{21}U^*_{\beta2}+ U_{\alpha1}aU^*_{\beta1}.
\label{Eq:51}
\end{align}
Since $\hat{H}_f$ is Hermitian, we have $H_{\alpha\beta}=H_{\beta\alpha}^*$.
\begin{center}
\begin{table}[ht]
\begin{tabular}{cc}
\toprule
%Mixing&Global fit\\
Parameters&Best fit$\pm1\sigma$\\
\hline
$\theta_{12}/^\circ$&$33.82^{+0.78}_{-0.76}$  \\
$\theta_{23}/^\circ$&$49.7^{+0.9}_{-1.1}$  \\
$\theta_{31}/^\circ$&$8.61^{+0.12}_{-0.13}$  \\
$\Delta m^2_{21}$&$7.39^{+0.21}_{-0.20}\times10^{-5}$eV$^2$\\
$\Delta m^2_{31}$&$2.525^{+0.033}_{-0.031}\times10^{-3}$eV$^2$\\
\toprule
\end{tabular}
\caption{Neutrino oscillation parameters (Best fit $\pm1\sigma$) \cite{Esteban2019}}
\label{Tab:1}
\end{table}
\end{center}
We then decompose the above flavour Hamiltonian by Hilbert-Schmidt decomposition given by Eq. (\ref{Eq:11}) and Eq. (\ref{Eq:12}) (See Appendix A). Given the flavour  Hamiltonian, we can numerically solve the Schr\"{o}dinger equation and find $\{\ket{\psi(t)}\}$. Then using Eq. (\ref{Eq:5}), we compute PBP for three flavour model using the global fit values given in TABLE (\ref{Tab:1}) and study its behaviour concerning $\delta_{CP}$ and $a$ for normal and inverted mass hierarchies.

\begin{figure}[ht]
	\includegraphics[width=1\columnwidth]{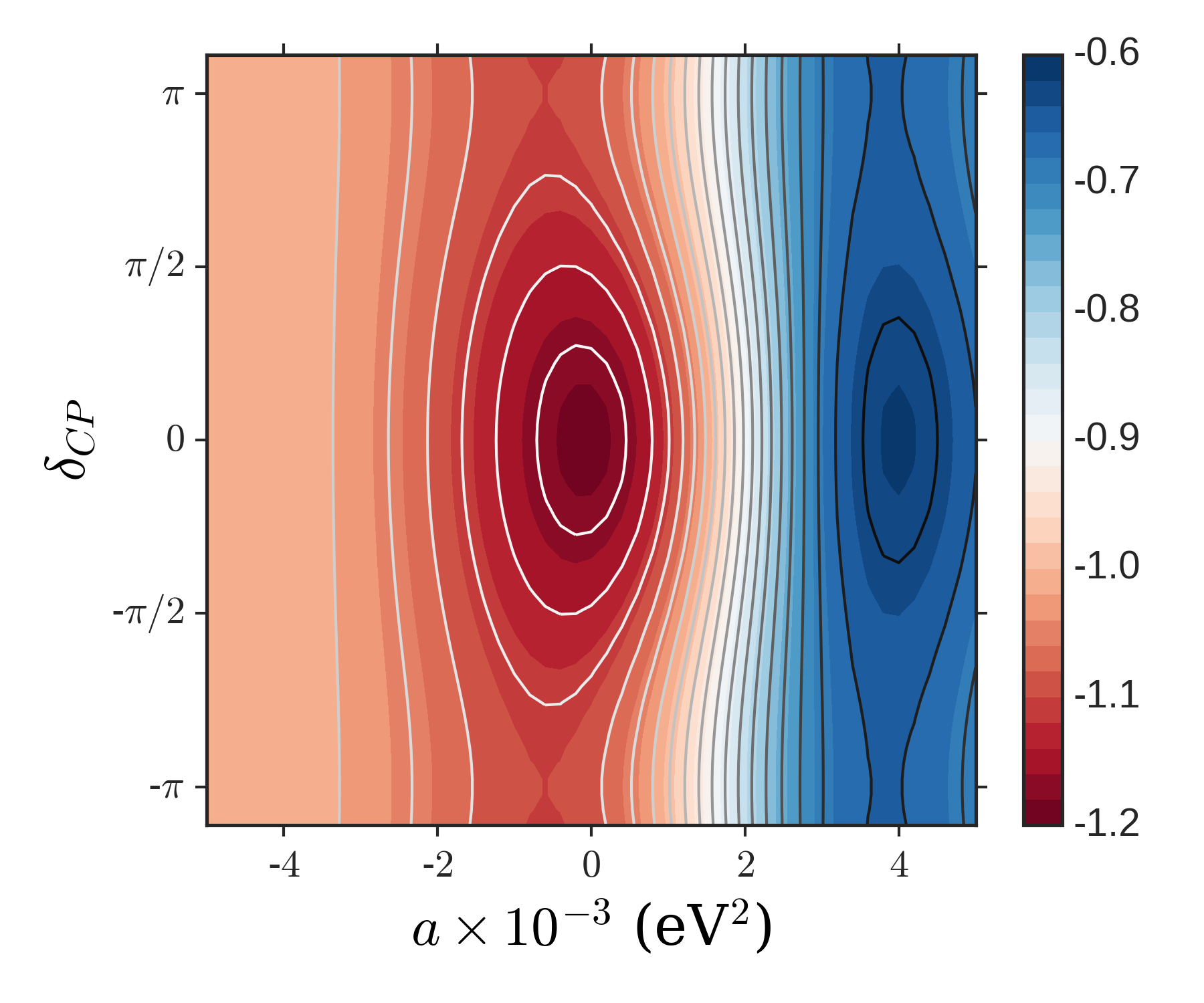}\\
	\caption{(Colour online) PBP plotted against $a$ and $\delta_{CP}$, using oscillation parameters given in TABLE (\ref{Tab:1}) with normal mass hierarchy.}
	\label{fig:10}
\end{figure}
\begin{figure}[ht]
	\includegraphics[width=1\columnwidth]{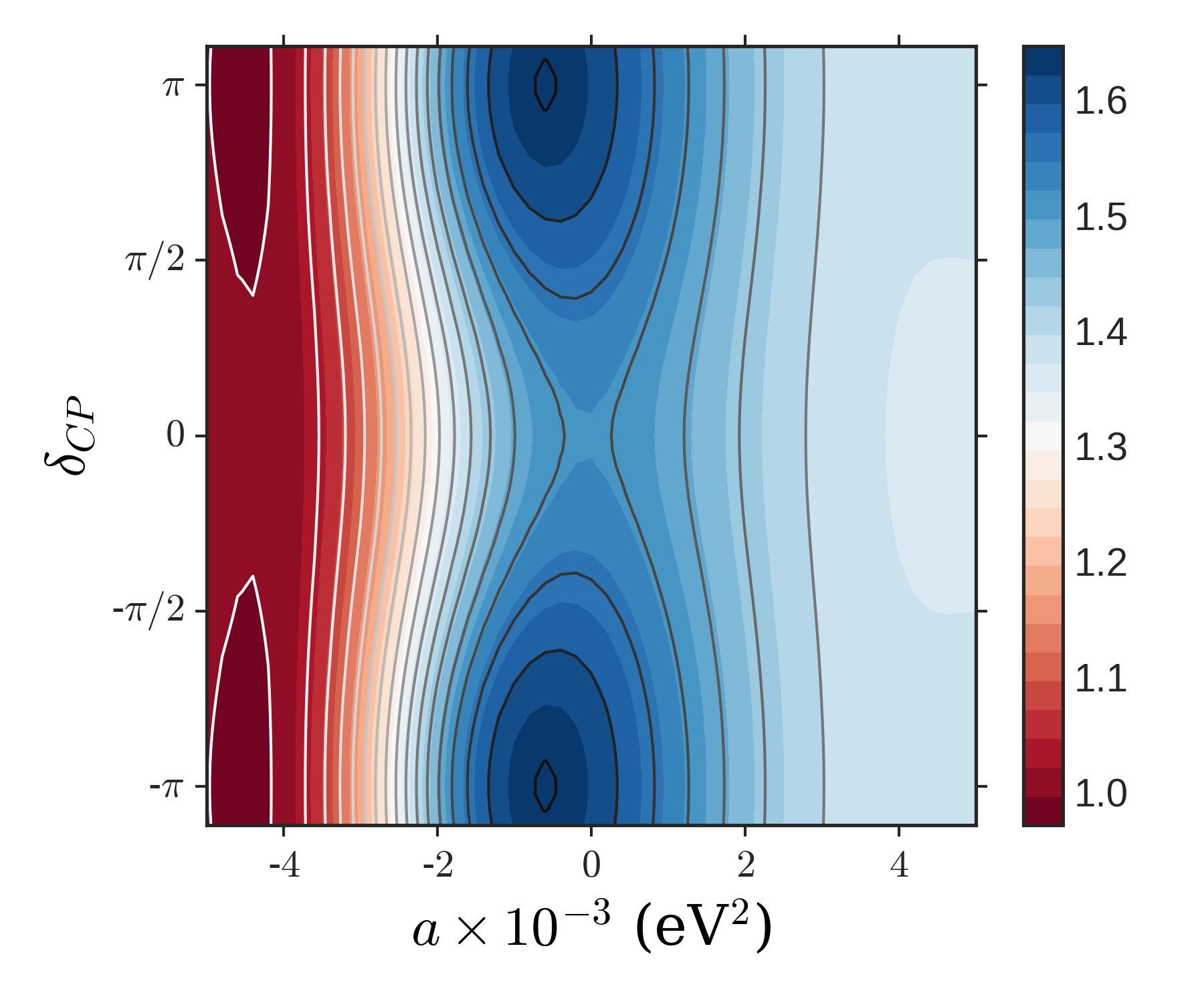}
	\caption{(Colour online) PBP plotted against $a$ and $\delta_{CP}$, using oscillation parameters given in TABLE (\ref{Tab:1}) with inverted mass hierarchy.}
	\label{fig:11}
\end{figure}
The results shown in Fig. (\ref{fig:10}) and Fig. (\ref{fig:11}) gives clear evidence for PBP being sensitive to mass hierarchies, $\delta_{CP}$ and matter potential. Direct estimation of PBP, given the type of neutrino and the matter potential, can estimate the magnitude of $|\delta_{CP}|$ and the mass hierarchy. The sign of $\delta_{CP}$ is a freedom in constructing the PMNS matrix. A negative value of $\delta_{CP}$ for neutrinos corresponds to a positive value of $\delta_{CP}$ for antineutrinos, and vice-versa. Conventionally, $\delta_{CP}$ is defined between $[0,2\pi)$ for neutrinos \cite{RevModPhys.87.531}. Since PBP is calculated using Bargmann invariant, we observe a symmetry in PBP for the different sign of $\delta_{CP}$. In the above calculation we have kept $L=979.7$ km, $E=2$ GeV and the initial state as $\ket{\nu_{\mu}}=\ket{010}$. 

The direct measurement of PBP in neutrinos is still under debate. Some works suggest the calculation of PBP from oscillation probabilities. The electromagnetic interactions proposed for neutrinos at one loop in the extended standard model can give more insights into PBP, as PBP can depend on slowly varying external electromagnetic fields. These features enable us to investigate the behaviour of neutrinos in extreme environments, such as stellar interiors. More robust extensions of our analysis can be made by including QFT corrections.

\section{Conclusions}
To summarise, we studied the Pancharatnam-Berry phase in neutrinos using the kinematic approach introduced by Mukunda and Simon. The method provides a more manageable platform to compute the Pancharatnam-Berry phase using Bargmann invariant. We reformulated the flavour Hamiltonian into two and three-qubit systems to investigate the behaviour of two and three flavour neutrino models, respectively. Our construction allows one to simulate the dynamics of flavour Hamiltonian in a quantum circuit. The construction also incorporates the effect of matter potential and the Dirac $CP$ phase. We derived the formula for the Pancharatnam-Berry phase by calculating the total and dynamical phases independently. The separate formulas highlight the behaviour of the Pancharatnam-Berry phase concerning the neutrino flavour mixing parameters. By imposing the cyclic evolution condition, our results match Blasone \textit{et al.} \cite{BLASONE1999262}.

In the equation for the total phase in two flavour approximation, we can see the nodal points when the term $\cos(2\theta)$ changes the sign. At $\cos(2\theta)$ equals zero, the total phase is not defined, and the Pancharatnam-Berry phase undergoes a sudden jump of $\pi$. Many authors missed this signature feature of nodal points in neutrino mixing. The term ``$\cos(2\theta)$'' is interesting as it also appears in identifying the MSW resonance points. In ordinary matter, neutrinos feel a positive matter potential and the MSW resonance appears only for negative values of $\cos(2\theta)$. For antineutrinos, the same occurs for positive values of $\cos(2\theta)$. Thus the resonance points will be less than $\pi/4$ for neutrinos and greater than $\pi/4$ for antineutrinos. Our studies reveal the same behaviour for nodal points. Hence, the nodal points correspond to the MSW resonance points, giving physical significance to them. We also trace the origin of $2\pi$ jumps and the bifurcations observed in the Pancharatnam-Berry phase to the dynamical phase. 

The numerical value and the sign of $\Delta m^2$ can impart a notable difference in the Pancharatnam-Berry phase. Since experiments exclude bi-maximal mixing, a varying energy experiment can address the mass hierarchies and their numerical bounds. For bi-maximal mixing, we lose the contrast, and the values converge to $\pi$. 

The Dirac $CP$ phase and the mass hierarchies can also leave their impressions on the Pancharatnam-Berry phase. For a given matter potential, the Pancharatnam-Berry phase is symmetric for positive and negative values of the Dirac $CP$ phase. Since neutrinos and antineutrinos feel different matter potentials, the Pancharatnam-Berry phase is sensitive to the magnitude of $\delta_{CP}$. We also investigated the effect of different mass hierarchies in the same framework and saw noticeable differences in the behaviour of the Pancharatnam-Berry phase. 

The direct measurement of the Pancharatnam-Berry phase in neutrino is still an open problem. Nevertheless, the kinematic approach enabled us to examine hidden features of neutrino mixing. Additional research on the Pancharatnam-Berry phase can bring more profound insights into the underlying physics of flavour mixing and the MSW effect. One can extend the work to various areas, such as non-standard interactions \cite{Sarkar2021}, sterile neutrinos \cite{Chatla2018} and many more. Being a phase with a topological origin, the Pancharatnam-Berry phase can, in principle, give a more robust platform for future quantum technologies, which makes this research appealing for a wider audience.

\begin{acknowledgments}
MTM sincerely thank C. P. Jisha, Friedrich Schiller Universit\"{a}t Jena, for the fruitful discussions on the Pancharatnam-Berry phase, Athul R. T., CUSAT, for assistance on numerical simulations and Sarath N., CUSAT, for discussions on neutrino physics. MTM thank CSIR-JRF, Government of India, (Grant No: 09/239(0558)/2019-EMR-I) for the financial support. MTM, TKM and RBT gratefully acknowledge the financial support by the Department of Science and Technology, India.
\end{acknowledgments}

\appendix
\section{Three flavour Hamiltonian}

We compute the matrix elements of Eq. (\ref{Eq:49}) by substituting Eq. (\ref{Eq:48})  to Eq. (\ref{Eq:50}). The exact formulas are,
\begin{widetext}
	\begin{align}
	H_{\tau\tau}=&\Delta{m^2_{21}} \left(- \sin{\theta_{23}} \cos{\theta_{12}} - e^{-i \delta} \sin{\theta_{12}} \sin{\theta_{13}} \cos{\theta_{23}}\right) \left(- e^{i \delta} \sin{\theta_{12}} \sin{\theta_{13}} \cos{\theta_{23}} - \sin{\theta_{23}} \cos{\theta_{12}}\right) \nonumber\\&+ \Delta{m^2_{31}} \cos^{2}{\theta_{13}} \cos^{2}{\theta_{23}}\\
	H_{\tau\mu}=&  \Delta{m^2_{21}} \left(\cos{\theta_{12}} \cos{\theta_{23}} - e^{-i \delta} \sin{\theta_{12}} \sin{\theta_{13}} \sin{\theta_{23}}\right)\left(- e^{i \delta} \sin{\theta_{12}} \sin{\theta_{13}} \cos{\theta_{23}} - \sin{\theta_{23}} \cos{\theta_{12}}\right) \nonumber\\&+ \Delta{m^2_{31}} \sin{\theta_{23}} \cos^{2}{\theta_{13}} \cos{\theta_{23}}\\
	H_{\tau e}=&  \Delta{m^2_{21}} \left(- e^{i \delta} \sin{\theta_{12}} \sin{\theta_{13}} \cos{\theta_{23}} - \sin{\theta_{23}} \cos{\theta_{12}}\right) \sin{\theta_{12}} \cos{\theta_{13}} + \Delta{m^2_{31}} e^{i \delta} \sin{\theta_{13}} \cos{\theta_{13}} \cos{\theta_{23}}\\
	H_{\mu\tau}=&\Delta{m^2_{21}} \left(- \sin{\theta_{23}} \cos{\theta_{12}} - e^{- i \delta} \sin{\theta_{12}} \sin{\theta_{13}} \cos{\theta_{23}}\right) \left(- e^{ i \delta} \sin{\theta_{12}} \sin{\theta_{13}} \sin{\theta_{23}} + \cos{\theta_{12}} \cos{\theta_{23}}\right) \nonumber\\&+ \Delta{m^2_{31}} \sin{\theta_{23}} \cos^{2}{\theta_{13}} \cos{\theta_{23}}\\
	H_{\mu\mu}=&\Delta{m^2_{21}} \left(\cos{\theta_{12}} \cos{\theta_{23}} - e^{-i \delta} \sin{\theta_{12}} \sin{\theta_{13}} \sin{\theta_{23}}\right) \left(- e^{ i \delta} \sin{\theta_{12}} \sin{\theta_{13}} \sin{\theta_{23}} + \cos{\theta_{12}} \cos{\theta_{23}}\right) \nonumber\\&+ \Delta{m^2_{31}} \sin^{2}{\theta_{23}} \cos^{2}{\theta_{13}}\\
	H_{\mu e}=&\Delta{m^2_{21}} \left(- e^{ i \delta} \sin{\theta_{12}} \sin{\theta_{13}} \sin{\theta_{23}} + \cos{\theta_{12}} \cos{\theta_{23}}\right) \sin{\theta_{12}} \cos{\theta_{13}} + \Delta{m^2_{31}} e^{ i \delta} \sin{\theta_{13}} \sin{\theta_{23}} \cos{\theta_{13}}\\
	H_{e\tau}=& \Delta{m^2_{21}} \left(- \sin{\theta_{23}} \cos{\theta_{12}} - e^{- i \delta} \sin{\theta_{12}} \sin{\theta_{13}} \cos{\theta_{23}}\right) \sin{\theta_{12}} \cos{\theta_{13}}+ \Delta{m^2_{31}} e^{- i \delta} \sin{\theta_{13}} \cos{\theta_{13}} \cos{\theta_{23}}\\
	H_{e\mu}=& \Delta{m^2_{21}} \left(\cos{\theta_{12}} \cos{\theta_{23}} - e^{- i \delta} \sin{\theta_{12}} \sin{\theta_{13}} \sin{\theta_{23}}\right) \sin{\theta_{12}} \cos{\theta_{13}} + \Delta{m^2_{31}} e^{- i \delta} \sin{\theta_{13}} \sin{\theta_{23}} \cos{\theta_{13}}\\
	H_{ee}=&\Delta{m^2_{21}} \sin^{2}{\theta_{12}} \cos^{2}{\theta_{13}} + \Delta{m^2_{31}} \sin^{2}{\theta_{13}}
	\end{align}
\end{widetext}
It is clear from the above equations, and the fact that the Hamiltonian being Hermitian, we have, $H_{\alpha\beta}=H_{\beta\alpha}^*$. This Hamiltonian can be decomposed into linear combination of tensor products of Pauli matrices including identity. The decomposed flavour Hamiltonian is given as,
\begin{widetext}
	\begin{align}
	\hat{H}_f=&\frac{1.27\times2}{8E}\left[\right.
	(H_{\mu\mu} + H_{\tau\tau} + H_{ee})		( \mathbf{1} \otimes \mathbf{1}\otimes \mathbf{1} )+
	( H_{\mu\mu} -  H_{\tau\tau} +  H_{ee})	 	( \mathbf{1} \otimes \mathbf{1}\otimes \sigma_z )+
	( H_{\mu\tau} +  H_{\tau\mu}	) 	( \mathbf{1} \otimes \sigma_x\otimes \sigma_x )\nonumber\\&+
	( i H_{\mu\tau} -  i H_{\tau\mu}	) 	( \mathbf{1} \otimes \sigma_x\otimes \sigma_y )+
	(- i H_{\mu\tau} +  i H_{\tau\mu})	 	( \mathbf{1} \otimes \sigma_y\otimes \sigma_x )+
	( H_{\mu\tau} +  H_{\tau\mu}	 )	( \mathbf{1} \otimes \sigma_y\otimes \sigma_y )\nonumber\\&+
	(- H_{\mu\mu} +  H_{\tau\tau} +  H_{ee}	) 	( \mathbf{1} \otimes \sigma_z\otimes \mathbf{1} )+
	(- H_{\mu\mu} -  H_{\tau\tau} +  H_{ee})	 	( \mathbf{1} \otimes \sigma_z\otimes \sigma_z )+
	( H_{e\tau} +  H_{{\tau}e}	) 	( \sigma_x \otimes \mathbf{1}\otimes \sigma_x )\nonumber\\&+
	(i H_{e\tau} - i H_{{\tau}e})	 	( \sigma_x \otimes \mathbf{1}\otimes \sigma_y )+
	( H_{e\mu} +  H_{{\mu}e}	 )	( \sigma_x \otimes \sigma_x\otimes \mathbf{1} )+
	( H_{e\mu} +  H_{{\mu}e}	 )	( \sigma_x \otimes \sigma_x\otimes \sigma_z )\nonumber\\&+
	( i H_{e\mu} -  i H_{{\mu}e}	 )	( \sigma_x \otimes \sigma_y\otimes \mathbf{1} )+
	( i H_{e\mu} -  i H_{{\mu}e}	 )	( \sigma_x \otimes \sigma_y\otimes \sigma_z )+
	( H_{e\tau} +  H_{{\tau}e})	 	( \sigma_x \otimes \sigma_z\otimes \sigma_x )\nonumber\\&+
	( i H_{e\tau} -  i H_{{\tau}e})	 	( \sigma_x \otimes \sigma_z\otimes \sigma_y )+
	(- i H_{e\tau} +  i H_{{\tau}e}	 )	( \sigma_y \otimes \mathbf{1}\otimes \sigma_x )+
	( H_{e\tau} +  H_{{\tau}e})	 	( \sigma_y \otimes \mathbf{1}\otimes \sigma_y )\nonumber\\&+
	(- i H_{e\mu} +  i H_{{\mu}e})	 	( \sigma_y \otimes \sigma_x\otimes \mathbf{1} )+
	(- i H_{e\mu} +  i H_{{\mu}e})	 	( \sigma_y \otimes \sigma_x\otimes \sigma_z )+
	( H_{e\mu} +  H_{{\mu}e})	 	( \sigma_y \otimes \sigma_y\otimes \mathbf{1} )\nonumber\\&+
	( H_{e\mu} +  H_{{\mu}e}	 )	( \sigma_y \otimes \sigma_y\otimes \sigma_z )+
	(- i H_{e\tau} +  i H_{{\tau}e})	 	( \sigma_y \otimes \sigma_z\otimes \sigma_x )+
	( H_{e\tau} +  H_{{\tau}e}	) 	( \sigma_y \otimes \sigma_z\otimes \sigma_y )\nonumber\\&+
	( H_{\mu\mu} +  H_{\tau\tau} -  H_{ee})	 	( \sigma_z \otimes \mathbf{1}\otimes \mathbf{1} )+
	( H_{\mu\mu} -  H_{\tau\tau} -  H_{ee}	 )	( \sigma_z \otimes \mathbf{1}\otimes \sigma_z )+
	( H_{\mu\tau} +  H_{\tau\mu}	 )	( \sigma_z \otimes \sigma_x\otimes \sigma_x )\nonumber\\&+
	( i H_{\mu\tau} -  i H_{\tau\mu}	 )	( \sigma_z \otimes \sigma_x\otimes \sigma_y )+
	(- i H_{\mu\tau} +  i H_{\tau\mu})	 	( \sigma_z \otimes \sigma_y\otimes \sigma_x )+
	( H_{\mu\tau} +  H_{\tau\mu}	 )	( \sigma_z \otimes \sigma_y\otimes \sigma_y )\nonumber\\&+
	(- H_{\mu\mu} +  H_{\tau\tau} -  H_{ee}	)	( \sigma_z \otimes \sigma_z\otimes \mathbf{1} )+
	(- H_{\mu\mu} -  H_{\tau\tau} -  H_{ee}	)	( \sigma_z \otimes \sigma_z\otimes \sigma_z )\left.\right]
	\label{Eq:A10}
	\end{align}
\end{widetext}
The factor $8$ comes to the equation due to the normalization factor of $2^n=2^3=8$ in the decomposition.  For three flavour oscillation in matter, one can follow the exact same steps followed in the two flavour scenario and can obtain similar formulas. Or one can decompose the matter potential $\hat{V}$ into the tensor products of Pauli matrices and add to the above equation.

\bibliographystyle{apsrev4-1}
\bibliography{ref}

\end{document}